\documentclass[11pt,a4paper]{article}
\usepackage{epsfig}
\usepackage{amsfonts}
\usepackage{amsmath}
\usepackage{rotating}
\usepackage{cite}
\newlength{\dinwidth}
\newlength{\dinmargin}
\def\lsim{\mathrel{\rlap{\lower4pt\hbox{\hskip1pt$\sim$}}
    \raise2pt\hbox{$<$}}} %less than or approx. symbol
\def\gsim{\mathrel{\rlap{\lower4pt\hbox{\hskip1pt$\sim$}}
    \raise2pt\hbox{$>$}}} %greater than or approx. symbol

\newcommand{\qqbar}{\mbox{$\mathrm{q}\bar{\mathrm{q}}$}}
\newcommand{\Gev}       {\mbox{${\rm GeV}$}}

\newcommand{\ra}        {\mbox{$ \rightarrow $}}
\newcommand {\pom} {I\hspace{-0.2em}P}

\newcommand {\alphapom} {\mbox{$\alpha_{_{\pom}}$}}
\newcommand {\alphappom} {\mbox{$\alpha^\prime_{_{\pom}}$}}
\newcommand {\rhoz}     {\mbox{${\rho^0}$}}  
\newcommand {\phiz}     {\mbox{${\phi}$}}
\newcommand {\rzfzz}     {\mbox{${r^{04}_{00}}$}}
\newcommand {\rzfpz}     {\mbox{${r^{04}_{10}}$}}
\newcommand {\rzfpm}     {\mbox{${r^{04}_{1-1}}$}}   

\newcommand {\Mpp}     {\mbox{${M_{\pi \pi}}$}}
\newcommand {\Mppsq}     {\mbox{${M_{\pi \pi}^2}$}}
\newcommand {\Mkk}     {\mbox{${M_{KK}}$}}
\newcommand {\Mee}     {\mbox{${M_{ee}}$}}
\newcommand {\jpsi}     {\mbox{${J/\psi}$}}

\newcommand {\Mrho}     {\mbox{${M_{\rho}}$}}

\newcommand {\Mrhosq}     {\mbox{${M_{\rho}^2}$}}
\newcommand {\Grho}     {\mbox{${\Gamma_{\rho}}$}}
\newcommand {\rhotopp} {\mbox{${\rho^0 \to \pi^+\pi^-}$}}
\newcommand {\phitokk} {\mbox{${\phi \to K^+K^-}$}}
\newcommand {\jpsitoee} {\mbox{${J/\psi \to e^+e^-}$}}
\newcommand {\jpsitomm} {\mbox{${J/\psi \to \mu^+\mu^-}$}}
\newcommand {\Grhosq}     {\mbox{${\Gamma_{\rho}^2}$}}
\newcommand {\Gz}     {\mbox{${\Gamma_{0}}$}}
\newcommand {\fhel}     {\mbox{${\phi_{h}}$}}

\newcommand {\Thel}     {\mbox{${\theta_{h}}$}}

\newcommand {\dsdt}     {\mbox{${\mbox{d}\sigma/\mbox{d}t}$}}
\newcommand {\dst}      {\mbox{${\frac{\rm{d}\sigma}{\rm{d}\it{t}}}$}}

\def\ap#1#2#3   {{\em Ann. Phys. (NY)} {\bf#1} (#2) #3}   
\def\apj#1#2#3  {{\em Astrophys. J.} {\bf#1} (#2) #3} 
\def\apjl#1#2#3 {{\em Astrophys. J. Lett.} {\bf#1} (#2) #3}
\def\app#1#2#3  {{\em Acta. Phys. Pol.} {\bf#1} (#2) #3}
\def\ar#1#2#3   {{\em Ann. Rev. Nucl. Part. Sci.} {\bf#1} (#2) #3}
\def\cpc#1#2#3  {{\em Comp. Phys. Commun.} {\bf#1} (#2) #3}
\def\err#1#2#3  {{\it Erratum} {\bf#1} (#2) #3}
\def\ib#1#2#3   {{\it ibid.} {\bf#1} (#2) #3}
\def\jmp#1#2#3  {{\em J. Math. Phys.} {\bf#1} (#2) #3}
\def\ijmp#1#2#3 {{\em Int. J. Mod. Phys.} {\bf#1} (#2) #3}
\def\jetp#1#2#3 {{\em JETP Lett.} {\bf#1} (#2) #3}
\def\jpg#1#2#3  {{\em J. Phys. G.} {\bf#1} (#2) #3}
\def\mpl#1#2#3  {{\em Mod. Phys. Lett.} {\bf#1} (#2) #3}
\def\nat#1#2#3  {{\em Nature (London)} {\bf#1} (#2) #3}
\def\nc#1#2#3   {{\em Nuovo Cim.} {\bf#1} (#2) #3}
\def\nim#1#2#3  {{\em Nucl. Instr. Meth.} {\bf#1} (#2) #3}
\def\np#1#2#3   {{\em Nucl. Phys.} {\bf#1} (#2) #3}
\def\pcps#1#2#3 {{\em Proc. Cam. Phil. Soc.} {\bf#1} (#2) #3}
\def\pl#1#2#3   {{\em Phys. Lett.} {\bf#1} (#2) #3}
\def\prep#1#2#3 {{\em Phys. Rep.} {\bf#1} (#2) #3}
\def\prev#1#2#3 {{\em Phys. Rev.} {\bf#1} (#2) #3}
\def\prl#1#2#3  {{\em Phys. Rev. Lett.} {\bf#1} (#2) #3}
\def\prs#1#2#3  {{\em Proc. Roy. Soc.} {\bf#1} (#2) #3}
\def\ptp#1#2#3  {{\em Prog. Th. Phys.} {\bf#1} (#2) #3}
\def\ps#1#2#3   {{\em Physica Scripta} {\bf#1} (#2) #3}
\def\rmp#1#2#3  {{\em Rev. Mod. Phys.} {\bf#1} (#2) #3}
\def\rpp#1#2#3  {{\em Rep. Prog. Phys.} {\bf#1} (#2) #3}
\def\sjnp#1#2#3 {{\em Sov. J. Nucl. Phys.} {\bf#1} (#2) #3}
\def\shep#1#2#3 {{\em Surveys in High Energy Phys.} {\bf#1} (#2) #3}
\def\spj#1#2#3  {{\em Sov. Phys. JEPT} {\bf#1} (#2) #3}
\def\spu#1#2#3  {{\em Sov. Phys.-Usp.} {\bf#1} (#2) #3}
\def\zp#1#2#3   {{\em Zeit. Phys.} {\bf#1} (#2) #3}
\def\ejp#1#2#3  {{\em Eur. Phys. J. } {\bf#1} (#2) #3}

\begin{document}
%
%
%
%                          TITLE
%
%
%
\begin{titlepage}
\title{\bf Measurement of diffractive photoproduction of vector mesons at
large momentum transfer at HERA}

\author{\large{\rm ZEUS Collaboration}}

\date{}
\maketitle
\vspace{2cm}

\begin{abstract}

\noindent
Elastic and proton--dissociative photoproduction of $\rho^0$, $\phi$
and $J/\psi$ vector mesons ($\gamma p\rightarrow Vp$, $\gamma
p\rightarrow VN$, respectively) have been measured in $e^+p$
interactions at HERA up to $-t=3$~GeV$^2$, where $t$ is the
four-momentum transfer squared at the photon--vector meson vertex.
The analysis is based on a data sample in which photoproduction
reactions were tagged by detection of the scattered positron in a
special-purpose calorimeter. This limits the photon virtuality, $Q^2$,
to values less than 0.01~GeV$^2$, and selects a $\gamma p$ average
center-of-mass energy of $\langle W\rangle$ = 94 GeV.  Results for the
differential cross sections, \dsdt, for \rhoz, \phiz\ and \jpsi\ 
mesons are presented and compared to the results of recent QCD
calculations.  Results are also presented for the $t$-dependence of
the pion-pair invariant-mass distribution in the \rhoz\ mass region
and of the spin-density matrix elements determined from the
decay-angle distributions.  The Pomeron trajectory has been derived
from measurements of the $W$ dependence of the elastic differential
cross sections \dsdt\ for both \rhoz\ and \phiz\ mesons.
\end{abstract}
\vspace{-16cm}
\begin{flushleft}
\tt DESY 99-160 \\
October 1999 \\
\end{flushleft}

\setcounter{page}{0}
\thispagestyle{empty}
\eject

\end{titlepage} 

\clearpage

%%%%% Here the author list ends.

\def\3{\ss}                                                                                        
\newcommand{\address}{ }                                                                           
\pagenumbering{Roman}                                                                              
                                    % this "%"s are for cosmetics only                             
%\begin{document}                                                                                   
                                                   %                                               
\begin{center}                                                                                     
{                      \Large  The ZEUS Collaboration              }                               
\end{center}                                                                                       
  J.~Breitweg,                                                                                     
  S.~Chekanov,                                                                                     
  M.~Derrick,                                                                                      
  D.~Krakauer,                                                                                     
  S.~Magill,                                                                                       
  B.~Musgrave,                                                                                     
  A.~Pellegrino,                                                                                   
  J.~Repond,                                                                                       
  R.~Stanek,                                                                                       
  R.~Yoshida\\                                                                                     
 {\it Argonne National Laboratory, Argonne, IL, USA}~$^{p}$
% ...
\newpage                                                                                           
% ...
$^{\    2}$ also at IROE Florence, Italy \\                                                        
% ...
                                                           %                                       
                                                           %                                       
% \par         % if index listing & table fit to 1 page, put gap here                              
\newpage   % alternatively: go to newpage, if page is too small                                    
                                                           %                                       
% \institute_references_start    % do not touch or move this line !                                
                                                           %                                       
\begin{tabular}[h]{rp{14cm}}                                                                       
$^{a}$ &  supported by the Natural Sciences and Engineering Research                               
          Council of Canada (NSERC)  \\                                                            
$^{b}$ &  supported by the FCAR of Qu\'ebec, Canada  \\                                            
$^{c}$ &  supported by the German Federal Ministry for Education and                               
          Science, Research and Technology (BMBF), under contract                                  
          numbers 057BN19P, 057FR19P, 057HH19P, 057HH29P, 057SI75I \\                              
$^{d}$ &  supported by the MINERVA Gesellschaft f\"ur Forschung GmbH, the                          
German Israeli Foundation, and by the Israel Ministry of Science \\                                
$^{e}$ &  supported by the German-Israeli Foundation, the Israel Science                           
          Foundation, the U.S.-Israel Binational Science Foundation, and by                        
          the Israel Ministry of Science \\                                                        
$^{f}$ &  supported by the Italian National Institute for Nuclear Physics                          
          (INFN) \\                                                                                
$^{g}$ &  supported by the Japanese Ministry of Education, Science and                             
          Culture (the Monbusho) and its grants for Scientific Research \\                         
$^{h}$ &  supported by the Korean Ministry of Education and Korea Science                          
          and Engineering Foundation  \\                                                           
$^{i}$ &  supported by the Netherlands Foundation for Research on                                  
          Matter (FOM) \\                                                                          
$^{j}$ &  supported by the Polish State Committee for Scientific Research,                         
          grant No. 115/E-343/SPUB/P03/154/98, 2P03B03216, 2P03B04616,                             
          2P03B10412, 2P03B03517, and by the German Federal                                        
          Ministry of Education and Science, Research and Technology (BMBF) \\                     
$^{k}$ &  supported by the Polish State Committee for Scientific                                   
          Research (grant No. 2P03B08614 and 2P03B06116) \\                                        
$^{l}$ &  partially supported by the German Federal Ministry for                                   
          Education and Science, Research and Technology (BMBF)  \\                                
$^{m}$ &  supported by the Fund for Fundamental Research of Russian Ministry                       
          for Science and Edu\-cation and by the German Federal Ministry for                       
          Education and Science, Research and Technology (BMBF) \\                                 
$^{n}$ &  supported by the Spanish Ministry of Education                                           
          and Science through funds provided by CICYT \\                                           
$^{o}$ &  supported by the Particle Physics and                                                    
          Astronomy Research Council \\                                                            
$^{p}$ &  supported by the US Department of Energy \\                                              
$^{q}$ &  supported by the US National Science Foundation                                          
\end{tabular}                                                                                      
                                                           %                                       
% \institute_references_end     % do not touch or move this line !                                 
                                                           %                                       

\newpage
\pagenumbering{arabic}
\setcounter{page}{1}
\normalsize
%
%
%%******************************************************************
%                   INTRODUCTION
%******************************************************************
%
\section{Introduction}

The study of exclusive diffractive $ep$ reactions at HERA has shown
that when $Q^2$ (photon virtuality) or the mass scale involved is  
large, the cross section increases with energy faster than expected 
for soft processes~\cite{jim,abramowicz-caldwell}.
The rise is consistent with predictions from models based on perturbative
QCD (pQCD) in which $Q^2$ and mass are used as the perturbative scale.
It is expected that, for diffractive
vector-meson photoproduction, the four-momentum transfer squared, $t$,
between the photon and the final-state vector meson may also serve as a hard
scale, provided $-t$ is large~\cite{large-t}.
In the present paper, this hypothesis is studied by
measuring diffractive vector-meson photoproduction as a function of $t$ 
and by confronting the data with the predictions of models based on pQCD, 
which should be applicable in the presence of a hard scale. In addition,
the data at lower $-t$ are compared to predictions of 
models expected to be valid for soft
processes. This provides a means to study the transition between the
soft, non-perturbative, and the hard, perturbative, regimes of
QCD~\cite{halfms}.
A detailed study of the onset of the pQCD regime should give 
important insight into the structure of strong interactions at hard scales
as well as improve our understanding of soft phenomena in QCD. 

Vector mesons can be diffractively photoproduced via two
processes. In one of them, the target proton
remains intact and the reaction is called exclusive (or elastic),
\begin{equation}
\gamma p \to V p.
\label{eq:el}
\end{equation}
In the other process, the proton dissociates into a higher mass
nucleonic state $N$ and the reaction is called proton-dissociative,
\begin{equation}
\gamma p \to V N.
\label{eq:pd}
\end{equation}
Reaction~(\ref{eq:el}) is called elastic in the framework of the
vector-dominance model (VDM)~\cite{vdm}, in which the photon
fluctuates into a virtual vector meson which in turn scatters
elastically from the target proton~\cite{bauer}. This reaction has
been studied over a wide range of $\gamma p$ center-of-mass energies
$W<200$~GeV~\cite{jim,bauer}.  The proton-dissociative
reaction~(\ref{eq:pd}) has been studied at low energies~\cite{chapin}
and information at high $W$ has been obtained recently~\cite{zrho-94}.

In this study, reactions~(\ref{eq:el})
and~(\ref{eq:pd}) are investigated with the ZEUS detector at HERA 
by measuring the processes $ e p \to e V p$ and 
$e p \to e V N$, 
where $V = \rho^0, \phi$ or $J/\psi$, and $N$ is a system with 
mass~$\le 7$~GeV.
The scattered positron was
detected in an electromagnetic calorimeter close to the beamline
at a distance of 44~m  from the interaction point in the direction 
of the outgoing positron. This ensured that the virtuality of the exchanged 
photon is very small ($Q^2 \le$ 0.01 GeV$^2$) 
and that $-t$ can be well approximated by the transverse momentum 
squared of the vector meson.

In the present paper, the pQCD 
\cite{bfgms,ryskin,bartels,ivanov,czerniak,ginzburg} 
and Regge \cite{collins} based approaches to
vector-meson production are described. Then the measurements of  
elastic and proton-dissociative photoproduction of the
$\rho^0, \phi$ and $J/\psi$ mesons are presented. Finally, the results
are compared with  models and a summary of the conclusions is given.

\section{Models}
\subsection{The pQCD approach -- hard scale models}
%%%%%%%%%%%%%%%%%%%%%%%%%%%%%%%%%%%%%%%%%%%%%%%%%%%%%%%%%%%%%%%%%
%%Here comes a description of the principles of pQCD models
%%%%%%%%%%%%%%%%%%%%%%%%%%%%%%%%%%%%%%%%%%%%%%%%%%%%%%%%%%%%%%%%%
In models based on pQCD~\cite{bfgms,ryskin,bartels,ivanov,czerniak,ginzburg},
the diffractive
photoproduction of a vector meson from a proton 
can be viewed 
in the proton rest frame
as a three-step process:
the photon fluctuates into a {\qqbar} state; the {\qqbar} pair
scatters on the proton target; and the scattered {\qqbar} pair becomes
a vector meson.  The {\qqbar} fluctuation is described in terms of
the photon wave-function derived from QCD.  The
interaction of the {\qqbar} pair with the proton is mediated in the
lowest order by the exchange of two gluons in a color-singlet state.
In the leading logarithmic approximation (LLA), the process can also 
be described
by the exchange of a gluon ladder \cite{ryskin,bartels,ivanov}.
To be calculable in pQCD, the interaction has to involve a hard
scale or, in other words, has to involve small transverse distances.
The transition of a {\qqbar} pair into a meson is, however, 
a non-perturbative phenomenon and can only be described in terms of the 
meson wave-function derived from lattice calculations and 
sum rules \cite{czerniak}.

The expected signatures of the perturbative regime in
diffractive meson production are:
\begin{itemize}
\item a fast rise of the diffractive cross sections with $W$,
the available center-of-mass energy, due
to the fast increase with decreasing $x$ (Bjorken scaling variable) 
of the gluon density in the proton \cite{bfgms,ryskin};
% YY: above suggested by Jim. W.
\item no variation with $W$ of the $t$-dependence of the cross section,
i.e. no $shrinkage$ of the diffractive peak \cite{bartels};
\item approximate restoration of flavor-independent production, which is
expected when the photon couples directly to the 
constituent quarks in the meson \cite{halfms};
\item production of light vector 
mesons in a helicity-zero state, independent of the 
initial photon helicity~\cite{ivanov,ginzburg}, in the asymptotic limit 
of very large $-t$ ($W^2\gg -t \gg \Lambda_{\rm{QCD}}^2$) .
\end{itemize}
%%%
\subsection{The Regge approach -- modeling the soft interactions}
%%%
Regge phenomenology~\cite{collins} has been successful in describing soft hadron-hadron  
interactions. In this approach, the interactions are described in terms of   
$t$-channel exchanges of Regge trajectories. In particular, diffractive   
processes are assumed to proceed through the exchange of the Pomeron
trajectory. 
%%%
\subsubsection{Regge factorization}
%%%
Regge factorization~\cite{collins} is the assumption that Regge pole residues 
factorize into a contribution from each vertex. In other
words, for diffractive vector-meson photoproduction, 
the properties of the interaction at the Pomeron-proton 
vertex should not depend on the properties of the Pomeron-vector-meson vertex. 
This hypothesis implies that
the ratio of elastic to proton-dissociative vector-meson photoproduction, 
$\dst (\gamma p \to V p)/\dst (\gamma p \to V N)$, should be the same for 
the three vector mesons under study in this paper.
In the framework of VDM, these ratios should have the same 
values as in hadron-proton reactions.
%%%
\subsubsection{The Pomeron trajectory}
%%%

In general, the differential cross section for a two-body hadronic
process, \dsdt , can be expressed at high energies as
\begin{equation}
\frac{d\sigma}{dt} = F(t) (W^2)^{[2\alphapom(t)-2]},
\label{eq:regge}
\end{equation}
where $F(t)$ is a function of $t$ only and $\alphapom(t)$ is the
Pomeron trajectory. At lower $W$ values, the exchange of a Reggeon trajectory
should also be taken into account. 

By studying the $W$ dependence of \dsdt\ at fixed $t$, $\alphapom(t)$
can be determined directly. Usually, the trajectory is assumed to be
linear, $\alphapom(t) = \alphapom(0) + \alphappom t$, but its form is
not predicted by Regge theory.  The early determinations of the
Pomeron trajectory according to this procedure used data from $p p$
elastic scattering~\cite{jl-slope,collins-slope}.

Under the assumption that \dsdt\ 
decreases exponentially (i.e. $\dsdt\sim\exp(bt)$), $\alphappom$
can also be determined from a study of the energy behavior 
of the exponential slope $b$. This method, however, is less direct  
and also depends on the $t$ range over which 
the exponent is fitted. Determinations of $\alphappom$ based on this 
procedure~\cite{giacomelli,burq-slope} from $ p p, K p$ and $\pi p$ elastic
scattering yielded values in the range of 0.14--0.28
GeV$^{-2}$. However, the high precision $p p$ ISR data at small $-t$ 
showed that $\alphappom$ has a value 0.25 GeV$^{-2}$ with a small
uncertainty~\cite{dl-slope}. The same analysis gave 
$\alphapom(0)=1.08$.

Studies of the elastic photoproduction of \rhoz\ and \phiz\ mesons 
have shown~\cite{jim} that these processes can be well described by the Regge
phenomenology developed for soft hadron-hadron collisions.
The steep energy behavior of the elastic $J/\psi$ photoproduction cross 
section at HERA, however, cannot be described in the Regge picture
by a Pomeron trajectory with an intercept of 1.08 but
requires a larger intercept. In addition, a direct
determination of the Pomeron trajectory in a way similar to that
described above has shown that the slope $\alphappom$ from elastic
photoproduction of $J/\psi$ is smaller than 0.25
GeV$^{-2}$~\cite{noshrink}. These observations suggest
that the Pomeron trajectory is not universal when a large
scale, like a large mass, is involved.  
It is thus of interest to see whether the
universality notion can be kept in soft interactions. Earlier attempts
to determine the Pomeron trajectory from the elastic photoproduction
of $\phi$ mesons~\cite{gl,barber-83} were not precise enough owing to the small
span in the energy available. The present measurements at HERA,
together with the existing lower-energy data, enable a more precise
determination of the Pomeron trajectory 
and thus make it possible to test its universality. 
 In this paper the Pomeron trajectory will be determined 
from the \rhoz\ and \phiz\ vector-meson data. The \jpsi\ measurement of this
analysis does not add significant information to the analysis done in 
\cite{noshrink} and therefore will not be considered here.

%******************************************************************
%                 EXPERIMENT 
%******************************************************************

\section{Experiment}
\label{setup}
The data used in the present analysis were collected with the ZEUS
detector at HERA in 1995, when HERA collided positrons of energy $E_e
= 27.5$ GeV with protons of energy $E_p =820$ GeV. The data sample used
in this analysis corresponds to an integrated luminosity of 1.98
pb$^{-1}$.

A detailed description of the ZEUS detector can be found
elsewhere~\cite{detector}.  A brief outline of the components  
which are most relevant for this analysis is given below.   

Charged particles are tracked by the central tracking detector (CTD),
which operates in a magnetic field of 1.43 T provided by a thin
superconducting coil. The CTD consists of 72 cylindrical drift chamber
layers, organized in 9 superlayers covering the polar 
angle\footnote{The ZEUS coordinates form a right-handed system
  with positive-$Z$ in the proton beam direction
  and a horizontal $X$-axis pointing towards the center of HERA.
  The nominal interaction point is at $X=Y=Z=0$.
  The polar angle $\theta$ is defined with respect to the $Z$
  direction.} region \mbox{$15^\circ < \theta < 164^\circ$.}
The transverse momentum 
resolution for 
full-length tracks
is 
$\sigma(p_t)/p_t=0.0058p_t\oplus 0.0065 \oplus 0.0014/p_t$,
with $p_t$ in GeV~\cite{ctd}.

The high resolution uranium-scintillator calorimeter (CAL) \cite{CAL}
consists of three parts: the forward (FCAL), the rear (RCAL) and the
barrel (BCAL) calorimeters.  Each part is subdivided transversely into
towers and longitudinally into one electromagnetic section (EMC) and
either one (in RCAL) or two (in BCAL and FCAL) hadronic sections (HAC).
The smallest subdivision of the calorimeter is called a cell.
 The CAL energy resolutions, as measured under
test beam conditions, are $\sigma(E)/E=0.18/\sqrt{E}$ for electrons and
$\sigma(E)/E=0.35/\sqrt{E}$ for hadrons ($E$ in GeV).

The proton-remnant tagger (PRT1) is used to tag events
in which the proton dissociates.  It consists of two layers of scintillation
counters perpendicular to the beam at $Z=5.15$~m.
The two layers are separated by a 2~mm thick lead absorber.
Each layer is split into two halves along the $Y$--axis and each half
is independently read out by a photomultiplier tube.  The counters
have an active area of $30\times 26$~cm$^2$
with a hole of $6.0\times 4.5$~cm$^2$ at the center to
accommodate the HERA beampipe.  The pseudorapidity range covered by
the PRT1 is \mbox{$4.3<\eta<5.8$}.

\begin{figure}[htb]
\begin{center}
\epsfig{file=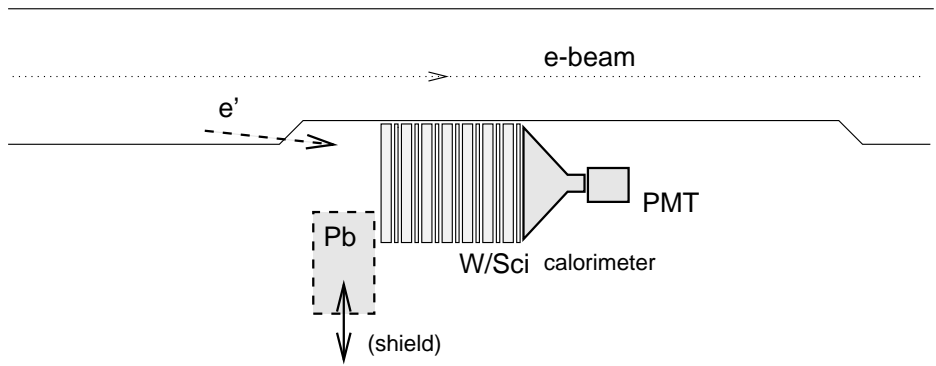,width=0.75\textwidth}
\end{center}
\vspace*{-0.5cm}
\caption{Sketch of the layout of the photoproduction tagging detector.}
\label{fig:tagger}
\end{figure}

The photoproduction tagger (PT) is a small 
electromagnetic calorimeter located at \mbox{$Z=-44$}~m, 
sensitive to 22--26~GeV positrons scattered under very small angles
(less than a few mrad). The HERA positron beampipe has a
14~mm deep and 60~cm long indentation on the side facing the ring
center (Fig.~\ref{fig:tagger}).  The calorimeter is installed behind a
1~mm thick copper window in the beampipe.
During beam injection 
and acceleration, a movable 10~cm thick lead shield is inserted
in front of the PT. The detector consists of twelve 
$70\times90\times7~\rm{mm}^3$ tungsten
plates interleaved with 3~mm thick scintillator layers. The
light from the scintillator is read out from the bottom by a
wavelength-shifter plate coupled through a plastic light-guide
to a photomultiplier tube with a quartz window (Philips XP1911).
The detector sensitive edge is about 28~mm from the positron
beam.

Additional scintillator strips are installed,
after each of the first three tungsten plates, at depths corresponding
to 2, 4 and 6 radiation lengths.  
These 8~mm wide vertical
strips are connected to plastic light-guides
coupled to three miniature photomultipliers (Hamamatsu R5600).  
Signals from the strips can be used to apply fiducial cuts and select 
well-contained electromagnetic showers.

Test beam measurements demonstrated that for 1--5 GeV electrons
hitting the calorimeter centrally, the energy resolution is 
$\sigma(E)/E = 0.25/\sqrt{E({\rm GeV})}$, 
and the calorimeter linearity is better than 1\%.
The energy measurement is used only at the trigger
level for tagging photoproduction events.
The tagger issues a trigger for events with an energy 
deposition above approximately 1~GeV.  The low threshold ensures 
that the tagging efficiency is determined mainly by the 
geometric acceptance.

The luminosity is determined from the rate of the Bethe-Heitler
bremsstrahlung process, $ep \rightarrow e \gamma p$, where the photon is 
measured with a calorimeter (LUMI) located in the HERA tunnel downstream of 
the interaction point in the direction of the outgoing positron~\cite{lumi}.
The acceptance of the LUMI calorimeter for Bethe-Heitler events is 
greater than 98\%.

%******************************************************************
%             KINEMATICS AND CROSS SECTIONS
%******************************************************************
\section{Kinematics and cross sections}
\label{Sect:Kinem}
\begin{figure}[htb]
\begin{center}
\vspace*{-0.5cm}
\epsfig{file=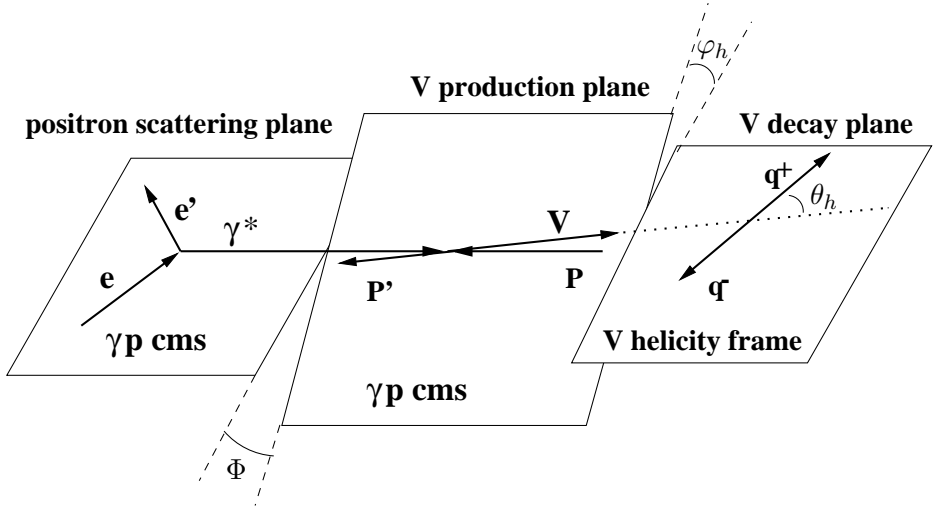,width=0.75\textwidth}
\put(-268,11){$\Phi$}
\put(-90,172){$\varphi_h$}
\put(-47,115){$\theta_h$}
\end{center}
\vspace*{-0.7cm}
\caption{Illustration of the angles used to analyze the helicity
states of the vector meson (for a decay into two particles, $V\rightarrow
\mbox{q}^+\mbox{q}^-$).}
\label{fig:hel_ang}
\end{figure}

The kinematics of the inclusive scattering of unpolarized 
positrons and protons are described by the squared positron-proton 
center-of-mass energy, $s$,
and any two of the following variables:
\begin{itemize}
    \item $Q^2$, the negative square
    of the exchanged photon's four-momentum;
    \item $y$, the fraction of the positron energy 
     transferred to the hadronic final state in the rest frame of the 
     initial-state proton;
    \item $W^2 = ys +M_p^2(1-y) - Q^2$, the squared
         center-of-mass energy of the photon-proton system (where
         $M_p$ is the proton mass); $ W^2 \approx ys$ in the case of
         photoproduction.
\end{itemize}
For the exclusive reaction $e p \rightarrow e V p$ ($V\rightarrow$
two charged particles) and the proton-dissociative process
$e p \rightarrow e V N$, $t$ and the following additional variables 
are used (see Fig.~\ref{fig:hel_ang}):
\begin{itemize}
    \item the angle, $\Phi$, between the $V$ production plane (which contains
   the momentum vectors of the virtual photon and the vector meson)
   and the positron scattering plane;
    \item the polar and azimuthal angles, $\theta_h$ and $\varphi_h$, of
         the positively-charged decay particle in the $V$ 
         helicity frame. The polar angle, $\theta_h$, 
is defined as the angle between the direction opposite to that of 
the outgoing proton and 
the direction of the positively-charged decay particle. 
The azimuthal angle, $\varphi_h$, is the angle between 
the decay plane and the $V$ production plane;
     \item the mass, $M_N$,
     of the diffractively-produced state $N$ in the proton-dissociative
     reaction.
     In the present analysis, however, it was not possible to measure this 
     quantity directly and the $M_N$ range covered was obtained from 
     Monte Carlo simulations (see Sect.~\ref{acc_corr}).
\end{itemize}
Only the three-momenta of the decay particles were
measured. Neither the momentum of the scattered positron, nor
the $\Phi$ angle, were measured. In such tagged
photoproduction events, $Q^2$ ranges from the kinematic
minimum, $Q^2_{\rm{min}} \approx M^2_e y^2/(1-y) \approx 
10^{-9}~\rm{GeV^2}$, where
$M_e$ is the electron mass, to a maximum value limited by the angular
acceptance of the PT,
$Q^2_{\rm{max}} \approx 
4E_eE_{e'}\sin^2(\theta_{\rm{max}}/2)\approx0.01~\rm{GeV^2}$, where 
$\theta_{\rm{max}}$ is the maximum scattering angle, and $E_e$ and $E_{e'}$ 
are the energies of the initial- and final-state positrons. 
The energy of the scattered positron is determined by the PT
acceptance. Since the typical $Q^2$ is very small (the median $Q^2$ is 
approximately $7 \times 10^{-6}~{\rm GeV^2}$), it can be
neglected in the reconstruction of the other kinematic variables.
The photon-proton center-of-mass energy is given by
\begin{equation}
    W^2 \approx 2 E_p (E - p_Z),
\label{Eq:W2Def}
\end{equation}
where $E_p$ is the incoming proton energy, and $E$ and
$p_Z$ are the energy and longitudinal momentum in the laboratory frame
of the produced meson $V$, respectively; in this approximation,
$M_N$, the meson mass, $M_V$, and its transverse momentum, 
$p_T$, are assumed to be much smaller than $W$.
The four-momentum transfer squared is given by
\begin{equation}
    t = (M_N^2-M_p^2-Q^2-M_V^2)^2/4W^2-(p_{\gamma}^*-p^*)^2 -4
p_{\gamma}^*p^*\sin^2(\theta^*/2) = 
 t_0-4 p_{\gamma}^*p^*\sin^2(\theta^*/2),
\label{Eq:tDef}
\end{equation}
where $p_{\gamma}^*$, $p^*$ are the magnitudes of the
photon and meson momenta and $\theta^*$ is the angle between them. The
starred quantities are defined in the
photon-proton center-of-mass system and $t_0$ is the
maximum $t(\theta^*=0)$ value.  For $Q^2\ll -t,p_T^2,M_V^2,M_N^2 \ll
W^2$, $t$ and $t_0$ are given by
\begin{equation}
    t \approx t_0 - p_T^2 \approx -M_V^2(M_N^2-M_p^2)/W^2 - p_T^2.
\label{Eq:ptrel}
\end{equation}
Since the maximum $-t_0$ value in the kinematic range covered by this
analysis is small ($-t_0 \lsim 7\times 10^{-3}~{\rm
GeV^2}$ for $V=\rhoz,\phiz$ and $-t_0 \lsim 6\times 10^{-2}~{\rm 
GeV^2}$ for $V=\jpsi$) compared to the $-t$ value, it can be neglected.
The four-momentum transfer squared  is then given by
\begin{equation}
    t \approx - p_T^2
\label{Eq:ptapr}
\end{equation}

The differential and integrated photoproduction cross sections for the
processes $\gamma p \rightarrow V p (N)$ were obtained from the
cross sections measured for the reactions $ep \rightarrow e V p
(N)$. The cross sections are related by
\begin{eqnarray}
\frac{d^2 \sigma_{ep}}{dydQ^2} &=& \frac{\alpha}{2\pi Q^2}
\left[ \frac{1+(1-y)^2}{y} - \frac{2(1-y)}{y} 
\frac{Q_{min}^2}{Q^2}\right]~ 
\sigma_{\gamma p}(W) \nonumber \\ 
&=&\varphi(y,Q^2)~\sigma_{\gamma p}(W),
\label{bornc}
\end{eqnarray}
where $\alpha$ is the fine structure constant and
$\varphi(y,Q^2)$ is the effective photon flux.
A measured $ep$  cross section can thus be transformed  into a
$\gamma p$ cross section,
\begin{equation}
\sigma_{\gamma p}= \frac{\sigma_{ep}}{\int \int 
\varphi(y,Q^2)\rm{d}\it{y}\rm{d}\it{Q^{\rm{2}}}}
 = \frac{\sigma_{ep}}{\Phi_\gamma},
\label{gp_cros}
\end{equation}
if $\sigma_{\gamma p}$ is independent of $W$ (or $y$) in the 
studied region (where $\Phi_\gamma$ is the integrated effective photon flux).
\section{Event selection}
\label{event_selection}
Vector mesons were observed in the two-body decay channels \rhotopp,
\phitokk, \jpsitoee, 
and \jpsitomm\ 
via the reconstruction of two oppositely-charged tracks in
the CTD. The scattered positron was detected in the PT
and the proton or its fragments escaped undetected or were
tagged in either the FCAL or the PRT1.

\subsection{Trigger}
\label{trigger}
ZEUS uses a three-level trigger system.  At the first level, a
coincidence between signals in the PT and a track candidate in the CTD
was required. Additionally, it was required that the energy deposition
in any of the FCAL towers closest to the beampipe should not exceed
1.25~GeV in order to suppress proton beam-related backgrounds. An
upper limit of 1~GeV on the energy deposited in the LUMI was also
imposed; this requirement suppressed events having a random
coincidence with bremsstrahlung.  The second and the third trigger
levels were mainly used to reject non-diffractive backgrounds by
requiring exactly two tracks pointing to the same vertex with a
$Z$-coordinate compatible with that of the nominal position of the
interaction point, $|V_{Z}| <\ 60$~cm.

The trigger efficiency was studied using 
data samples selected by two independent triggers
and was found to be about 90\% for the elastic events used in this
analysis and flat in all relevant kinematic variables. The
efficiency for the proton-dissociative events was significantly lower,
of the order of 10\%, due to the FCAL energy requirement.
This requirement restricts the mass of the dissociative 
system to $M_N\lsim 7$~GeV.

\subsection{Offline requirements}
\label{offline}
In the offline event selection the following conditions were imposed:
\begin{itemize}
\item exactly two oppositely-charged tracks from a common vertex;
\item each track with $p_t > 0.15 $~GeV  and
  $|\eta| < 2.2$; 
\item the vertex coordinates in the range 
$V_R\equiv\sqrt{V_X^2 + V_Y^2} < 0.7$ cm and $|V_Z| < 40$ cm;
\item  $85 < W < 105 \ \Gev $, thereby selecting a region of high
and well understood tagging efficiency;
\item a maximum energy of $200$ MeV (RCAL) and $250$ MeV (BCAL)
deposited in any
calorimeter cell, with the exception of those matched to tracks;
\item energy deposition in any of the 8 FCAL towers closest to the
beampipe less than 1.20 GeV.
\vspace*{-0.2cm}
\end{itemize} 
The number of events thus selected was 25446.

The final identification of the vector-meson candidates was performed
using cuts on the invariant mass of the track pairs measured in the CTD.
The pion, kaon and electron masses were in turn assigned to the
tracks, leading to the selection of  22823 \rhoz\ 
($ 0.55 <  \Mpp < 1.20 \ \Gev, \Mkk > 1.075 \ \Gev $), 366
\phiz\ ($ 0.99 < \Mkk < 1.06 \ \Gev $), and 120
\jpsi\ ($ 2.98 < \Mee < 3.13 \ \Gev $) meson candidates. 
The invariant mass resolution varies from 5 to 60~MeV depending on
the meson type and the values of $t$ and $W$. 
\vspace*{-0.3cm}
\section{Acceptance corrections}
\label{acc_corr}
\subsection{Monte Carlo generators}
The exclusive reaction $e p \rightarrow e V p$ was modeled using
the DIPSI~\cite{dipsi} Monte Carlo (MC) program.
For the simulation of the reaction $ep \rightarrow e  V N$, 
the EPSOFT Monte Carlo generator~\cite{epsoft} was used. The $M_N$ 
distribution at fixed $t$ was reweighted to the dependence
\begin{equation}
\frac{\rm{d}\sigma_{\gamma p \rightarrow VN}}{\rm{d} \it{M_{N}}^2} \propto
\frac{1}{M_N^{\beta(t)}}.
\label{e:beta}
\end{equation}
For \rhoz\ and \phiz\ production, the function
$\beta(t)=1.12e^{0.6t+0.3t^2+0.04t^3}+1.08e^{0.85t+0.11t^2}$ was 
used;
this function was found by parameterizing the ISR \cite{chlm} 
data on $\frac{\rm{d}\sigma}{\rm{d} \it{M_{N}^2}}$ 
in $pp$ single diffraction at large $-t$ and $M_N^2<0.1W^2$~\cite{LA_thesis}.
For \jpsi\ production the average between the above $\beta(t)$ and 
$\beta(t)=2.35$ (expected in Regge phenomenology for constant 
$\alphapom(t) = 1.175$~\cite{jpsi_intercept}) was used.
 
The $t$ (and $M_{\pi\pi}$ for $V$=\rhoz ) distribution was reweighted 
in both generators so
as to reproduce the measured distribution after reconstruction; 
the polar and azimuthal angular distributions of the decay
particles in the helicity frame were also reweighted.

\subsection{Photoproduction tagging acceptance}
\begin{figure}[htb]
\vspace*{-1.cm}
\begin{center}
\epsfig{file=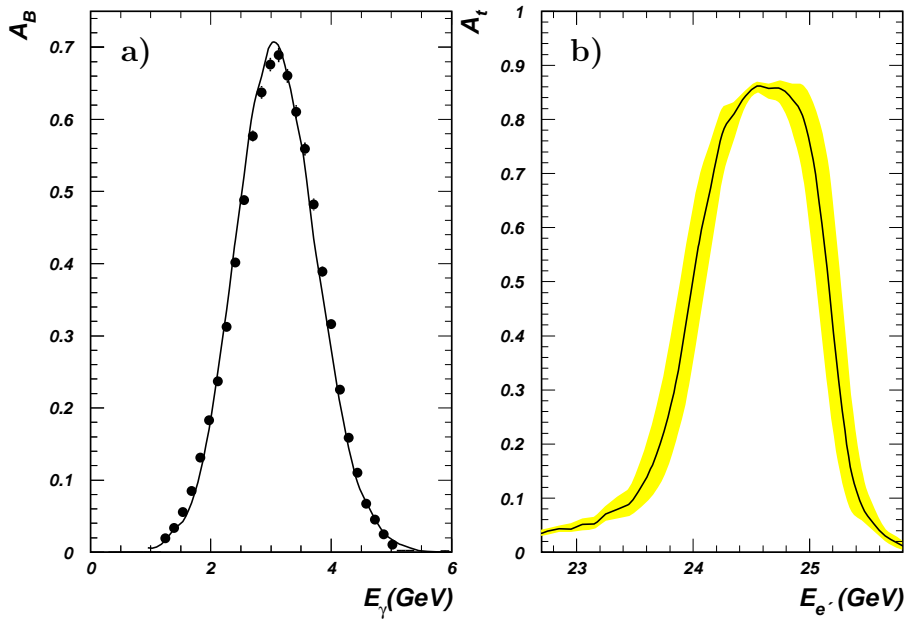,width=0.8\textwidth}
\put(-330,220){\bf \large a)}
\put(-160,220){\bf \large b)}
\end{center}
\vspace*{-1.0cm}
\caption{a) The measured bremsstrahlung tagging efficiency (dots),
$A_B$, as a 
function of the measured photon energy, $E_\gamma$, compared to the MC
expectation (solid line). (b) Photoproduction tagging efficiency $A_t$
as a function of
the energy of the scattered positron $E_e'$; the shaded band
represents the systematic uncertainty.}
\label{fig:tagger_acc}
\end{figure}

The geometric acceptance of the PT was simulated by a program which uses
the HERA beam-transport matrices to track the positron through the HERA
beamline. This program was tuned so as to reproduce the measured tagging
efficiency, $A_{B}$, using  Bethe-Heitler events, 
$e p \rightarrow e \gamma p$. $A_{B}$ was defined
as the fraction of bremsstrahlung events with a photon measured
in the LUMI photon detector when the PT
fired the trigger.  
In Fig.~\ref{fig:tagger_acc}a, $A_{B}$ is shown as a
function of the measured photon energy, $E_\gamma$, in the LUMI;
the MC prediction is in reasonable agreement with the data. 
The acceptance for photoproduction events was determined using the
geometric acceptance of the tagger and events generated according to
the equivalent-photon approximation for positron scattering angles up
to 3~mrad (for larger angles the acceptance is negligible); for these
events, the positron was tracked through the HERA beamline. The
photoproduction tagging efficiency, $A_t$, was calculated as a
function of the positron energy, $E_{e^\prime}$, (see
Fig.~\ref{fig:tagger_acc}b). For the kinematic range used in this
analysis, the average (cross-section weighted) PT acceptance was 70\%.
The systematic error was evaluated by changing the data sets used to
tune the MC and by varying the photon energy scale and the position of
the positron exit window in MC within their systematic uncertainties.
\subsection{Overall acceptance}
The generated events were processed  through the same chain 
of selection and reconstruction procedures as the data, thereby accounting 
for trigger as well as detector efficiencies (except for that of the PT) 
and smearing effects in the ZEUS detector. The reconstructed 
Monte Carlo events were then weighted with the function
$A_t(E_{e^\prime})$ in order to account for the PT acceptance and efficiency.

\begin{figure}[htb]
\vspace*{-.3cm}
\begin{center}
\epsfig{file=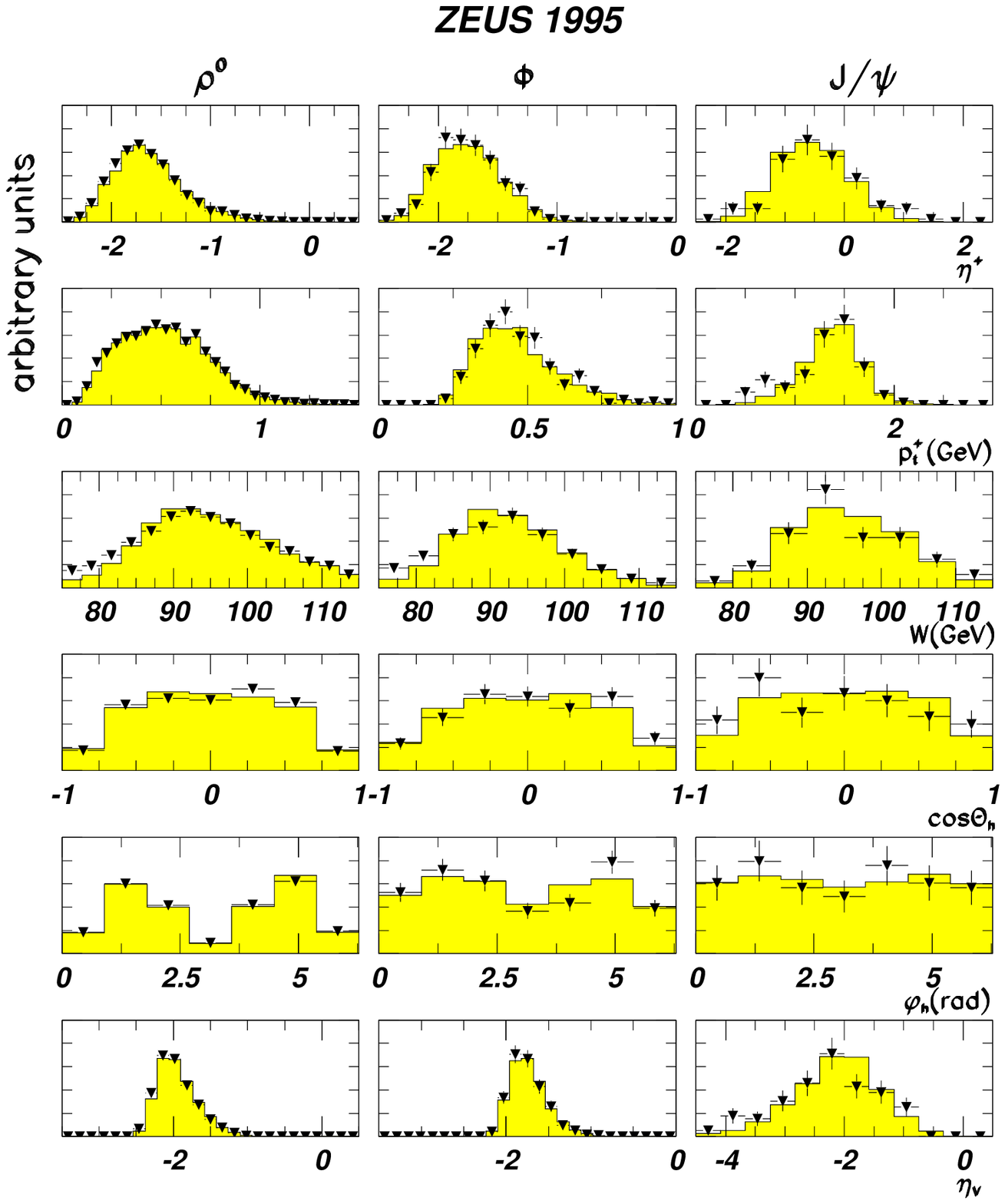,width=0.9\textwidth}
\end{center}
\vspace*{-1.6cm}
\caption{Comparison between the data and MC distributions of 
$\eta$ and $p_t$ of the positively charged track, and $W$, 
$\cos\theta_h$, $\phi_h$ and the meson pseudorapidity 
$\eta_{V}$. The three columns refer to the sum of the elastic and 
proton-dissociative \rhoz, \phiz\ and \jpsi\ samples, respectively.} 
\label{fig:compmc}
\end{figure}

All measured
distributions are well described by the Monte Carlo simulations. Some
examples are displayed in Fig.~\ref{fig:compmc}.
The overall acceptance in a
given bin was then determined as the ratio of the number of accepted
Monte Carlo events (weighted by $A_t$)
to the number generated in the selected kinematic
range. The acceptance, calculated in this manner, accounts for the
geometric acceptance, the detector and reconstruction efficiencies,
the detector resolution and the trigger efficiency.

\begin{figure}[htb]
\begin{center}
\vspace*{+.5cm}
\epsfig{file=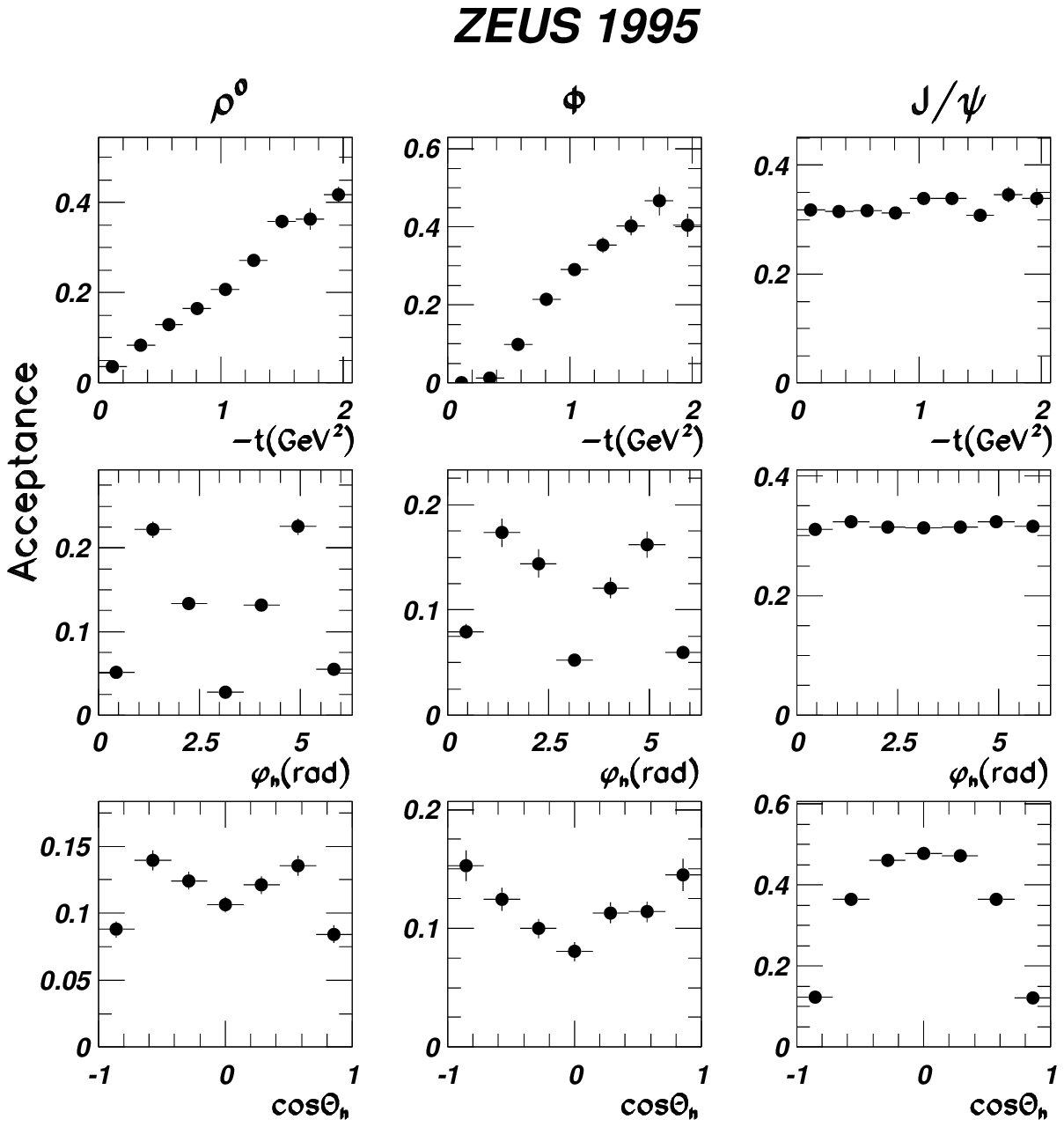,width=0.9\textwidth}
\end{center}
\caption{Overall acceptance for elastically  produced
\rhoz, \phiz\ and \jpsi\ mesons as 
a function of $-t$ and the helicity variables $\phi_h$ and $\cos\theta_h$.}
\label{fig:acceptance}
\end{figure}

Figure~\ref{fig:acceptance} shows the overall acceptance for elastic
events as a function of $t$, $\varphi_h$ and $\cos{\theta_h}$.
Inefficiencies at small $-t$, as well as the strong variation of the
acceptance with $\varphi_h$ for \rhoz\ and \phiz\ mesons, are mainly due to
the relatively small opening angle between the decay particles, resulting
in many very backward tracks (as can be deduced from the
pseudorapidity distributions of 
tracks in Fig.~\ref{fig:compmc}) outside the
geometric acceptance of the CTD.

The average acceptances are of the order of 30\% for
elastic and 5\% for proton dissociative events, respectively. The much lower
acceptance for the proton dissociative events is mainly due to the
FCAL energy requirement imposed in the trigger. 
%
% Backgrounds
%
\section{Backgrounds}
The dominant background sources are non-resonant $\pi^+\pi^-$ production
for the \rhoz\ analysis, \rhoz\ production for the \phiz\ analysis, and 
Bethe-Heitler $\gamma\gamma\rightarrow e^+e^-/\mu^+\mu^-$ 
production for the \jpsi\ case.
These backgrounds were statistically subtracted using
the fits to invariant-mass distributions as described
in Sect.~\ref{sec:results}.

The background due to inclusive photon diffractive dissociation,
$\gamma p\rightarrow Xp$, was studied using Monte Carlo simulations
and minimum-bias data samples. It was found to be about 1-2\% and was
also subtracted in the fitting procedure.

In the \rhoz\ analysis, the backgrounds due to decays of 
$\omega$ and $\phi$ mesons were found from MC studies to be negligible.
The background due to non-diffractive events was also small and
was neglected.

On average, 10\% of the diffractive events were rejected by the trigger
due to accidental coincidences with bremsstrahlung events in the LUMI
calorimeter. A correction was applied to account for this effect.
A small fraction of events, below 1\%, was selected due to fake
photoproduction tagging from bremsstrahlung overlays
(when the meson decay particles were measured by the CTD and the scattered 
positron was undetected, while the PT was hit instead by a
bremsstrahlung positron); this effect was neglected. 
\section{Separation of elastic and proton-dissociative processes}
\label{sec:sep}
The selected samples of diffractively-produced vector mesons are
mixtures of elastic and proton-dissociative events. These two processes 
have been separated as a function of $t$ on a statistical
basis using the MC simulation.

Proton-dissociative events were tagged by requiring a signal
in one of the PRT1 counters above a threshold corresponding to the
signal of a minimum ionizing particle. Alternatively, for systematic checks, 
energy depositions in the FCAL towers close to the beampipe were also used
for tagging the proton-dissociative events.

\begin{figure}[htb]
\begin{center}
\epsfig{file=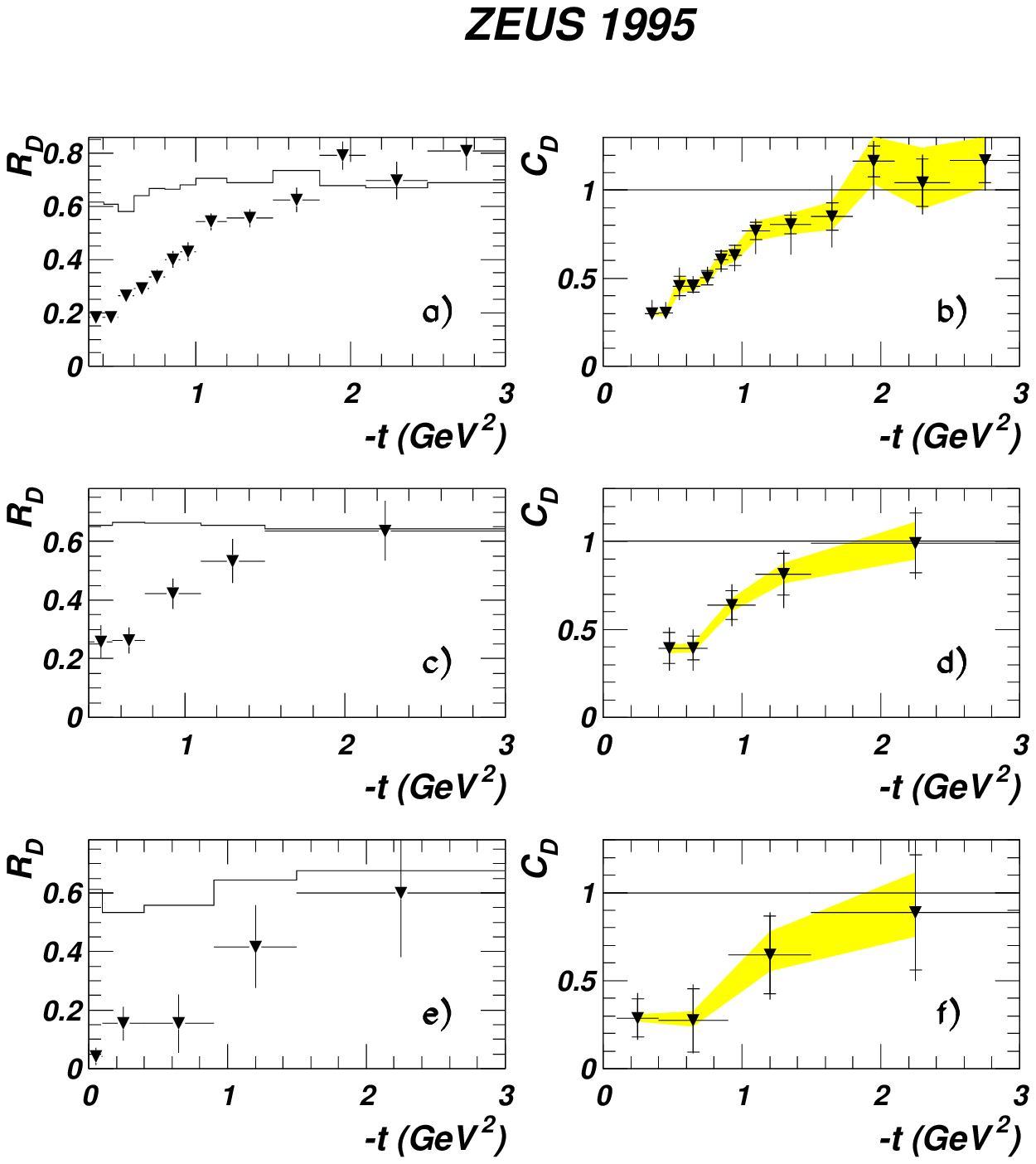,width=\textwidth}
\end{center}
\caption{Observed fraction of PRT1 tags, $R_{D}$, in the data (triangles
with statistical error bars) and in the proton-dissociative MC 
(histogram), and the estimated fraction of 
proton-dissociative events in data, $C_{D}$, for the 
\rhoz \ (a,b), \phiz \ (c,d) and \jpsi \ (e,f) samples as
a function of $-t$. The inner error bars indicate the statistical errors,
the outer bars the statistical and systematic uncertainties added in 
quadrature. The shaded bands in b), d) and f) represent the size of the 
correlated errors due to modeling of the proton dissociation in the
Monte Carlo.}
\label{fig:sep}
\end{figure}

Since the non-diffractive backgrounds were negligible, the following
relation was assumed:
$$\frac{N^T_{pd,data}}{N_{pd,data}} = \frac{N_{pd,MC}^T}{N_{pd,MC}},$$
where $N_{pd,data}$ and $N_{pd,MC}$ are the numbers of all 
accepted (but not necessarily tagged)
proton-dissociative events, whereas $N_{pd,data}^T$ and $N_{pd,MC}^T$ 
are the numbers of tagged proton-dissociative events, 
in the data and proton-dissociative EPSOFT samples, respectively. 
Therefore, the fraction, $C_D$, of proton-dissociative events in 
the data was calculated from
$$ C_D=\frac{N_{pd,data}}{N_{data}}=\frac{N^T_{pd,data}}{N_{data}}/
\frac{N_{pd,MC}^T}{N_{pd,MC}},$$
where $N_{data}$ is the number of all observed events (elastic and
proton-dissociative) in a given $t$ bin in the data.
In Fig.~\ref{fig:sep} the observed fraction of PRT1 tags, $R_{D}$, 
in the data and in the proton-dissociative MC,
as well as the estimated fraction of proton-dissociative events,
$C_D$, are displayed separately for the 
$\rho^0$, $\phi$ and $J/\psi$ samples. 
For the low $-t$ region, the
elastic process contributes a large fraction of the selected
diffractive events. For $-t>1.0$ GeV$^2$, the contribution from the
proton-dissociative process exceeds that from the elastic 
channel.

\clearpage

\section{Systematic uncertainties}
The systematic uncertainties  were subdivided 
into those related to the selection procedure and the 
detector simulation, and those reflecting 
the uncertainty in the model used for
the Monte Carlo generator. 

The effects of the following changes in the
selection procedure were checked:
\renewcommand{\theenumi}{\arabic{enumi}}
\renewcommand{\labelenumi}{\theenumi)}
\begin{enumerate}
\item 
the FCAL (instead of PRT1) was used to tag the
proton-dissociative events; 
\item 
the matching procedure between the tracks and energy depositions in CAL was
varied by changing the matching distance between the track and CAL cluster and
the energy requirement used to define the CAL cluster;
\item 
the minimum value of the track transverse momentum was
varied between 100 and 200 MeV;
\item 
the limit on the track $|\eta|$ was varied between 2.1 and 2.3;
\item the requirements on the vertex position were varied by amount
corresponding approximately to the resolution, i.e. $\pm$ 5 cm in $Z$
and $\pm$ 0.25 cm in the radial direction;
\item
the selected invariant mass region was changed by widening and narrowing it
by amounts corresponding to approximately the appropriate mass resolution;
\item
the threshold values of all
PRT1 counters were increased by 100\%. 
\end{enumerate}

The first check resulted in 5--20\% changes of the measured cross sections.
Checks 2--5 resulted in 3--5\% changes, and the last two checks had  
negligible effect. 

In the MC simulation, the positions of some beamline elements (eg. the
position of the 
synchrotron collimator jaws) were varied within their uncertainties, 
resulting in a 5--15\% change in the cross section. 

In order to estimate the systematic uncertainty from the uncertainties in
the parameters assumed in the EPSOFT generator, the following
modifications were made:
\begin{itemize} 
\item 
the shape of $\beta(t)$ was varied within the uncertainties of its
measurement~\cite{LA_thesis}; 
for \rhoz\ and \phiz\ production, this corresponds to  
$\beta(t)=2.24e^{0.6t+0.3t^2+0.04t^3}$ and 
$\beta(t)=2.16e^{0.85t+0.11t^2}$; for \jpsi\ production, 
to 
$\beta(t)=1.12e^{0.6t+0.3t^2+0.04t^3}+1.08e^{0.85t+0.11t^2}$
and $\beta(t)=2.35$. These variations significantly changed
the acceptance corrections, by up to 10\% at low $-t$ and up to 70\% for
the proton-dissociative sample at the highest $-t$;
\item
multiplicity distributions of the decay particles of the
dissociative system $N$ were varied within the uncertainties
of their measurement~\cite{cool}. This resulted in cross section
variations of $<5\%$.
\end{itemize}
Additionally, the re-weighting of other MC distributions (of the decay
particle angles or invariant masses, for example) was performed
in the range allowed by maintaining satisfactory agreement between data
and Monte Carlo. The effect on the cross sections was $<3\%$. 
In the determination of the \rhoz\ spin-density matrix elements,
the difference between the nominal method (see Sect. \ref{sec:hel}) and the
method of moments was taken as an additional uncertainty.

The overall normalization error due to the photoproduction 
tagging uncertainty was $\pm15$\%.
\section{Results}
\label{sec:results}
The differential cross sections \dsdt\ for elastic and 
proton-dissociative photoproduction of $V$ 
were evaluated in each bin of $t$ as:
\begin{equation}
\label{e:dsdt_el}
\frac{\rm{d}\sigma_{\gamma p \ra V p}}{\rm{d}\it{t}} = 
\frac{N\cdot (1-C_{D})\cdot C_{res}}
{A\cdot {\cal L}\cdot \Phi_\gamma \cdot \Delta t \cdot C_{br}},
\end{equation}

\begin{equation}
\label{e:dsdt_pd}
\frac{\rm{d}\sigma_{\gamma p \ra V N}}{\rm{d}\it{t}} = \frac{N\cdot C_{D}\cdot
C_{res}}{A\cdot {\cal L}\cdot \Phi_\gamma \cdot \Delta t \cdot C_{br}},
\end{equation}
 
where $N$ is the number of observed vector-meson candidates in bin
$\Delta t$ after all selection cuts, $C_{D}$ is the estimated fraction
of the proton-dissociative events in the bin, $C_{res}$ is the resonant
contribution in the bin, $C_{br}$ is the branching ratio of the
vector-meson decay mode considered, $A$ is the overall acceptance in the 
bin, ${\cal L}$ the integrated luminosity, and $\Phi_\gamma$ is the integrated 
effective photon flux (see Eq.~\ref{gp_cros}). In the kinematic region
$Q^2 < (E_e\theta_{max})^2(1-y)$, where $\theta_{max}=$~3~mrad
(see Sect. 6.2), and $85< W < 105$~GeV, $\Phi_\gamma=0.0121$.
The branching ratios for the \rhotopp,
\phitokk, \jpsitoee\ or \jpsitomm\ decay modes
were taken as $ 1,\ 0.5$ and $0.12$, respectively.
Effects due to QED radiation,
estimated to be smaller than 2\%~\cite{kurek}, were neglected.

A similar procedure was used to evaluate the differential cross section
$\rm{d}\sigma/\rm{d}\Mpp$.

For the proton-dissociative reaction, cross sections are extrapolated to 
$M_N^2=0.1W^2$ using the EPSOFT MC, as modified by Eq.~\ref{e:beta}.
\subsection{$\rho^0$ photoproduction }

\subsubsection{Resonance mass shape}
\label{sec:mass}

The differential cross sections $d\sigma/d\Mpp$ were fitted using a 
parameterization inspired by the S\"oding model~\cite{soding},
where the p-wave relativistic Breit-Wigner (BW) shape is distorted
by the interference with non-resonant $\pi\pi$ production:
\begin{equation}
\frac{d\sigma}{d\Mpp} = A^2 \left[
\left | \frac{ \sqrt{\Mpp \Mrho \Grho}}{\Mppsq-
\Mrhosq+i\Mrho\Grho}+B/A \right | ^2 + f \right], 
\label{soding}
\end{equation}
where \Mrho \ is the \rhoz \ mass, \Grho \ is the momentum-dependent
 width
\begin{equation}
\Grho(\Mpp) = \Gz\left( \frac{q}{q_0} \right)^3 \frac{\Mrho}{\Mpp},
\label{width}
\end{equation}
\Gz \ is the width of the \rhoz , $q$ is the $\pi$ momentum in
the $\pi\pi$ rest frame and $q_0$ is the value of $q$ at $M_{\pi\pi}
= M_{\rho}$.  The non-resonant amplitude (taken to be 
$M_{\pi\pi}$-independent) is
denoted by $B$, and $A$ is the normalization factor of the resonant
amplitude.  Additionally, another term, $f$, was introduced to account for
the background from reactions with photon diffractive
dissociation. The~term $f$ was assumed to be linear in $M_{\pi\pi}$,
$f\propto(1+1.5M_{\pi\pi})$~\cite{zrho-94} with $M_{\pi\pi}$ in GeV.
Alternatively the following parameterization, proposed by Ross and 
Stodolsky~\cite{rs}, was used:
\begin{equation}
\frac{d\sigma}{d\Mpp} = A\left[
\frac{ \Mpp \Mrho \Grho}{(\Mppsq-
\Mrhosq)^2+\Mrhosq\Grhosq} \left(\frac{\Mrho}{\Mpp} \right)^{n} + f \right] ,
\label{ross}
\end{equation}

where \Grho \ is given by Eq.~\ref{width} and $f$ has the same
form as described above.

\begin{figure}[htb]
\vspace*{-0.5cm}
\begin{center}
\epsfig{file=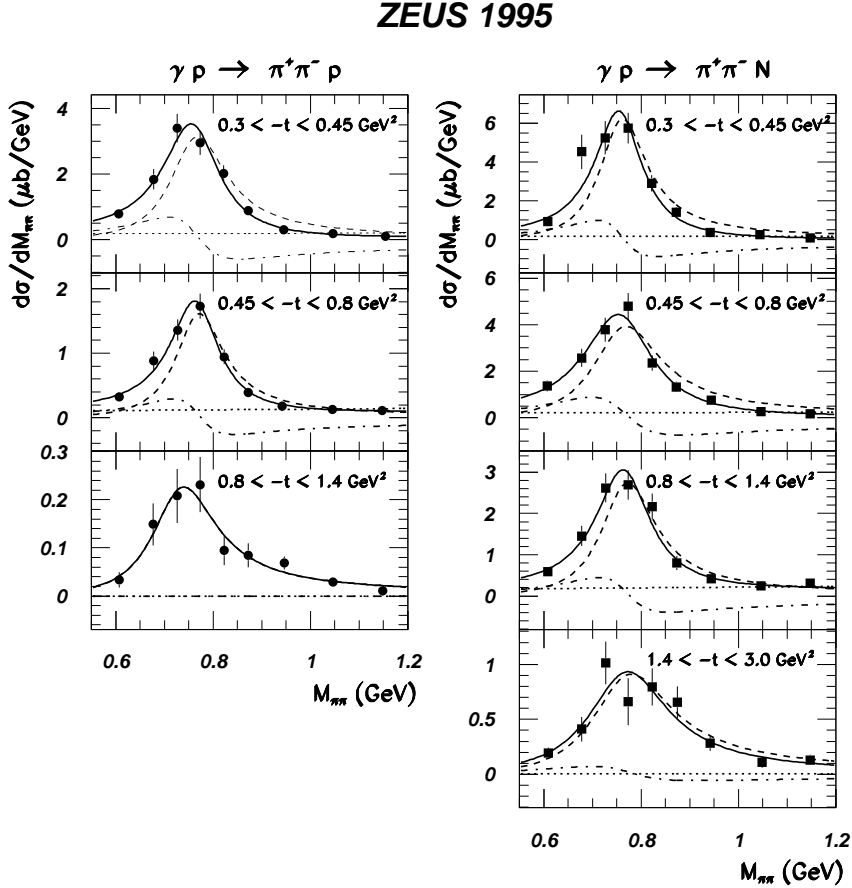,width=0.8\textwidth}
\end{center}
\vspace*{-1cm}
\caption{The differential cross sections $d\sigma/d\Mpp$ for 
several $t$ ranges. The
points represent the data and the curves indicate the result of the fits
with  Eq.~\ref{soding}. The dashed curves represent the
resonant contribution, the dotted curves the non-resonant contribution and 
the dot-dashed curves the contribution from the interference term. The 
solid curves are the sum. Only the statistical errors are shown. The 
circles and squares correspond to the elastic and the
proton-dissociative samples, respectively.}
\label{fig:rho_mas_t}
\end{figure}

The $\chi^2/NDF$ for all the fits is satisfactory. The fitted 
values of \Mrho,
\Gz \ and $f$ do not depend on the prescription used to parameterize
the mass distribution. The mass, \Mrho, and width, \Gz, for all $t$
values are
compatible with the values of the Particle Data Group (PDG)\cite{ref:pdg}. 
The level of background under the \rhoz \ peak, as given by the
integral of the function $f$, is about 1--2\%.  The mass
distributions for elastic and proton-dissociative \rhoz \ production
are shown in Fig.~\ref{fig:rho_mas_t} 
 together with the results of
the fits using Eq.~\ref{soding}. The mass resolution varies
 between 20 and 60~MeV for \Mpp\ between 0.5 and 1.2 GeV,
respectively. 

\begin{figure}[htb]
\begin{center}
\epsfig{file=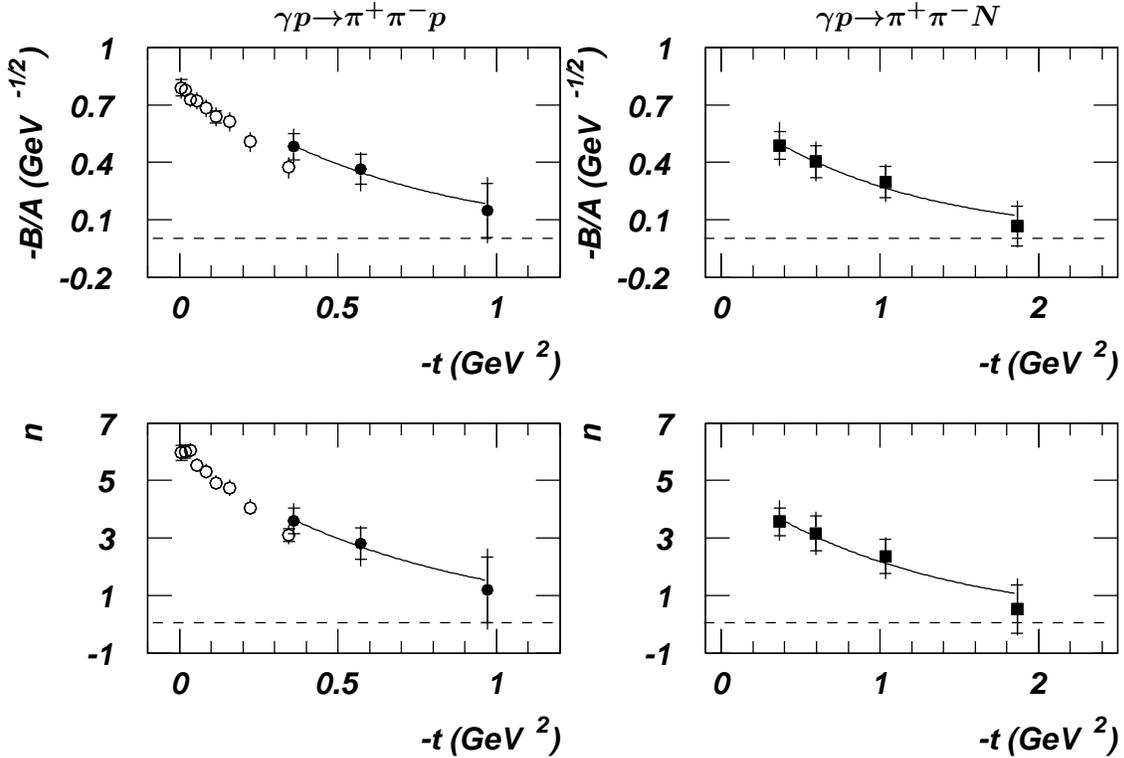,width=0.9\textwidth}
\put(-320,275){\boldmath $\gamma p \ra \pi^+ \pi^- p$}
\put(-115,275){\boldmath $\gamma p \ra \pi^+ \pi^- N$}
\end{center}
\vspace*{-.2cm}
\caption{The ratio $B/A$ from Eq.~\ref{soding} and the
parameter $n$ from Eq.~\ref{ross}, as a function of $-t$.  
The inner error
bars indicate the statistical errors, the outer ones the statistical and
systematic uncertainties added in quadrature. The solid circles and
squares correspond to the elastic and the
proton-dissociative samples, respectively. The 
open circles correspond to the low $-t$ ZEUS results \cite{zrho-94}. 
The solid curves are fits of the form $k e^{b_{S}t}$ (see text).}
\label{fig:mas_res}
\end{figure}

The fits to the mass distributions were repeated with 
\Mrho \ and \Gz \ fixed to the PDG values.  
The results for $B/A$ and $n$ are shown as a function of $-t$ in
Fig.~\ref{fig:mas_res}; they match with our earlier 
measurement for the low $-t$ region at
$\langle W\rangle=70$~GeV~\cite{zrho-94}. 
Both $B/A$ and $n$ decrease with $-t$, indicating
that the resonance shape distortion decreases.
This decrease is much faster
with $-t$ than that with $Q^2$~\cite{zvm95}.

The dependence of the ratio $B/A$ on $t$ was parameterized as
\begin{equation}
B/A = k e^{b_{S}t}.
\label{n_t}
\end{equation}
The fit to the data yields $k = -0.86\pm
0.52$~(stat.)~GeV$^{-1/2}$ and $b_{S} = 1.6 \pm 0.9$~
(stat.)~GeV$^{-2}$ for the elastic reaction, and $k = -0.71 \pm 0.21$~
(stat.)~GeV$^{-1/2}$ and $b_{S} = 0.9 \pm 0.4$~(stat.)~GeV$^{-2}$ for
the proton-dissociative process. The similar magnitudes
and $t$ dependences of the
$B/A$ ratio for the elastic and the proton-dissociative processes
indicate that the data are consistent with the hypothesis of
factorization of the diffractive vertices.

The \rhoz\ production cross section was evaluated in the mass
range $2M_\pi<\Mpp<M_\rho+5\Gamma_0$. The extrapolation from
the measured range $0.55<\Mpp<1.2~\Gev$ was made using the 
results of the fit to Eq.~\ref{soding}.

%
% dsigma/dt
%
\subsubsection{$t-$distributions}

\begin{figure}[htb]
\vspace*{-1.0cm}
\begin{center}
\epsfig{file=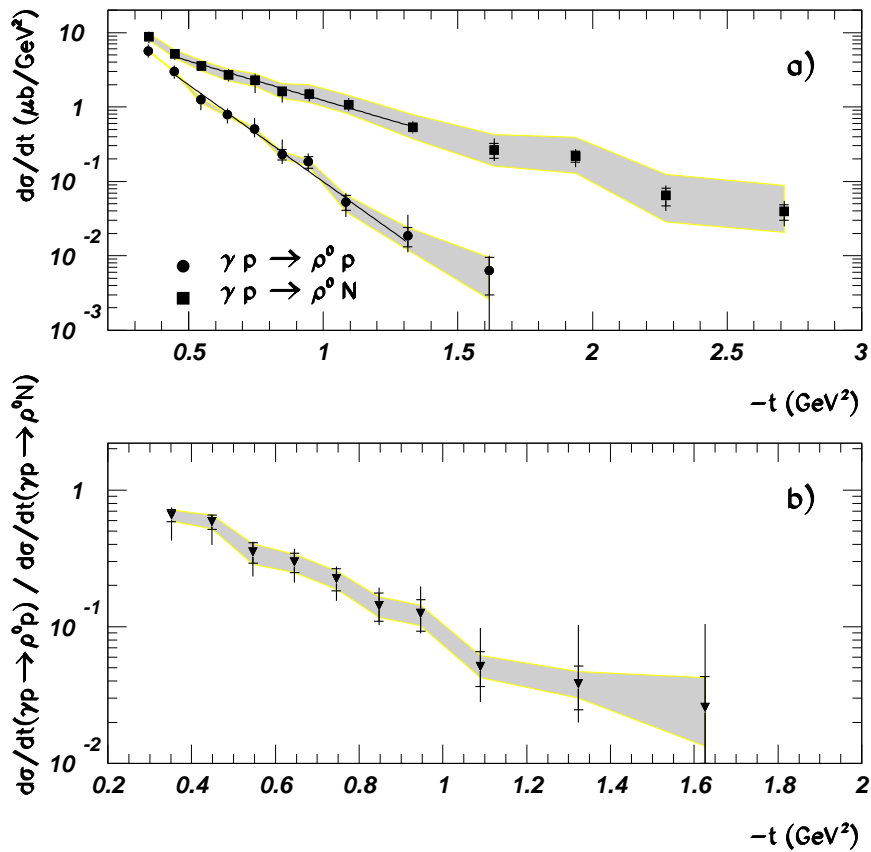,width=0.81\textwidth}
\end{center}
\vspace*{-1.0cm}
\caption{a) The differential cross sections \dsdt for elastic
(circles) and proton-dissociative (squares) \rhoz \
photoproduction.
The solid lines represent the results of the fit with the
function $Ae^{bt}$. The normalization error of 15\% is not shown.
b) The ratio of the elastic to the proton-dissociative cross
sections shown in a). The inner error bars indicate the statistical errors,
the outer bars the statistical and systematic uncertainties added in 
quadrature. 
The shaded bands represent the size of the correlated errors due
to the modeling of the proton dissociation in the 
Monte Carlo. 
}
\label{fig:dsdt-rho}
\end{figure}

The differential cross sections, $\dsdt$, for elastic and
proton-dissociative (for $M^2_N<0.1W^2$) 
$\rho^0$ photoproduction are plotted in
Fig.~\ref{fig:dsdt-rho}a.  Both samples exhibit an exponential
drop with increasing $-t$, with the cross section for the elastic 
process falling off more steeply than that for 
the proton-dissociative process. A fit with the function 
$A\exp(bt)$ in the range $0.4<-t<1.2$~GeV$^2$ for the elastic process 
gives $b_{el}$ = 6.0 $\pm 0.3~(\mbox{stat.})~^{+0.6}_{-0.3}~(\mbox{sys.}) 
\pm 0.4~(\mbox{mod.})$~GeV$^{-2}$ and
$b_{pd}$ = 2.4 $\pm 0.2~(\mbox{stat.})~^{+0.2}_{-0.1}~(\mbox{sys.})\pm 
0.3~(\mbox{mod.})$ GeV$^{-2}$  for the proton-dissociative sample,
where (mod.) represents the uncertainty due to the modeling 
of the proton dissociation in the EPSOFT Monte Carlo. 

The ratio of the elastic to the
proton-dissociative cross sections, $\rm{d}\sigma(\gamma p \to \rho^0
p)/ \rm{d}\sigma(\gamma p\to \rho^0 N)$, in a given $t$ interval is
shown in Fig.~\ref{fig:dsdt-rho}b.
This ratio falls off rapidly with $-t$ from about unity at $-t\approx$
0.4 GeV$^2$ to about $10^{-2}$ for $-t\ge$ 1 GeV$^2$. By comparison,
this ratio has the value 2.0 $\pm$ 0.2~(stat.) $\pm$ 0.7~(syst.) for
$-t<$ 0.5 GeV$^2$ at $W$ = 70 GeV \cite{zrho-94}.  An exponential fit
to the ratio gives a value for the slope $\Delta b=3.4
\pm0.3$~(stat.)~$^{+1.5}_{-0.3}$~(sys.)~ $\pm 0.2$~(mod.)~GeV$^{-2}$,
which is similar to that found previously \cite{zrho-94}.

\begin{figure}[htb]
\begin{center}
\epsfig{file=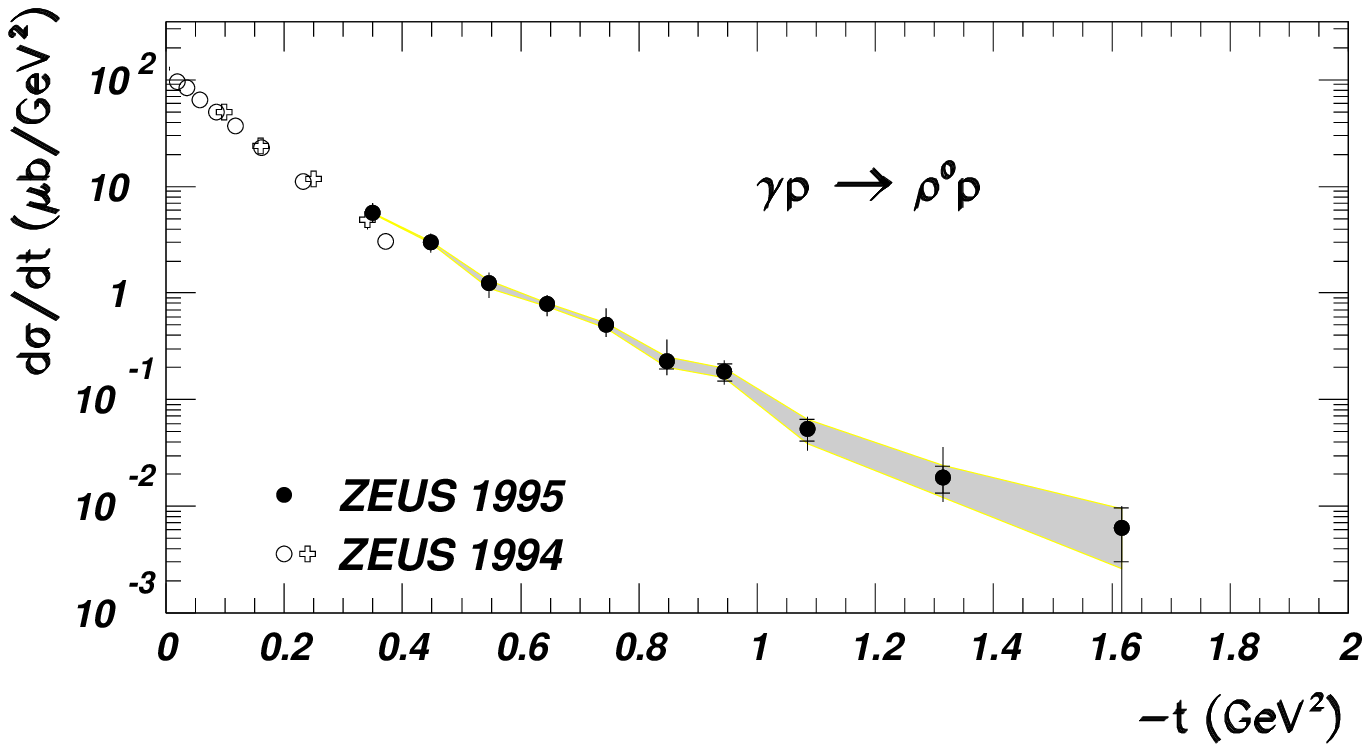,width=0.95\textwidth}
\end{center}
\vspace*{-1.0cm}
\caption{The differential cross sections, \dsdt, for elastic
\rhoz \ photoproduction. The solid circles are those shown in 
Fig.~\ref{fig:dsdt-rho}; the open circles and crosses are from 
earlier ZEUS data~\cite{zrho-94,zrho-lps}. 
The shaded band represents the size of the correlated errors due
to the modeling of the 
proton dissociation in the Monte Carlo. The normalization error of
15\% is not shown.
%The solid
%curve represent the results of the combined fit with the function $d\sigma
%/d\abst = Ae^{-b\abst+ct^2} $, where $b=11.1\pm 0.2$ and $c=3.2\pm 0.3$ 
}
\label{fig:dsdt-rho-el}
\end{figure}

Figure~\ref{fig:dsdt-rho-el} shows the differential cross
section for the elastic reaction together with results obtained 
in our earlier studies of the
low $-t$ region~\cite{zrho-94,zrho-lps}. 
The data at  low $-t$
were measured at a somewhat lower $W$ ($\langle W\rangle=70$ GeV) and have been
rescaled to the average $W$ of the present analysis 
($\langle W\rangle=94$~GeV),
assuming the Pomeron trajectory as measured in this analysis 
(see Sect.~\ref{trajectory}). 
The large $-t$ data match well with the low $-t$ results.

The data in Fig.~\ref{fig:dsdt-rho-el} cannot be described
by the expression $A\exp(bt)$ 
over the whole $t$ range. The slope $b_{el}$ 
(of about $10~\Gev^2$~\cite{zrho-94})
in the low $-t$ region is larger than that in the 
large $-t$ region, in contrast to the case for $\Delta b$,
which is the same within errors
in both $t$ regions, as well as at higher $Q^2$~\cite{h1-rho-deltab}.

%
% Helicity stuff
%
\subsubsection{Decay angular distributions}
\label{sec:hel} 
The angular distributions of the decay pions were used to determine
some of the $\rho^0$ spin-density matrix elements using the invariant
mass selection $0.55<\Mpp<1.2~\Gev$. The direction of the virtual
photon was approximated by that of the incoming positron.  In this
measurement, the three-dimensional angular distribution has been
averaged over the azimuthal angle between the positron scattering
plane and the $\rho^0$ production plane, $\Phi$, and thus no longer
distinguishes the photon helicity states $\pm$ 1.  The normalized
two-dimensional decay angular distribution can be written as
\cite{wolf}
\begin{eqnarray}
W(\cos{\Thel},\fhel) = & \frac{3}{4\pi} \left\{ \frac{1}{2}(1-\rzfzz)+
\frac{1}{2}(3\rzfzz - 1)\cos^{2}{\Thel} \right. \nonumber \\ 
                      & - \sqrt{2} \mbox{Re}[\rzfpz]\sin{2\Thel}\cos{\fhel}
                       \left. - \rzfpm\sin^{2}{\Thel}\cos{2\fhel} \right\} ,
\label{hel_2d_dis}
\end{eqnarray}
where the spin-density matrix element \rzfzz \ represents the probability
that the produced \rhoz \ has helicity $0$; the element \rzfpz \ is
related to the interference between the helicity non-flip and single-flip 
amplitudes; 
\rzfpm \ is related to the interference between the non-flip and 
double-flip amplitudes.  If $s$-channel helicity conservation 
(SCHC)~\cite{schc}
holds, \rzfzz, \rzfpm \ and $\mbox{Re}[\rzfpz]$ \ should be zero.

The parameters \rzfzz, \rzfpm \ and Re$[\rzfpz]$  were obtained by
minimizing the difference between the two-dimensional  ($\cos{\Thel},
\fhel$) angular distributions of the data and those of the simulated
events, which were re-weighted according to Eq.~\ref{hel_2d_dis}. 
A binned $\chi^2$-method was used.

\begin{figure}[htb]
\vspace*{-.5cm}
\begin{center}
\epsfig{file=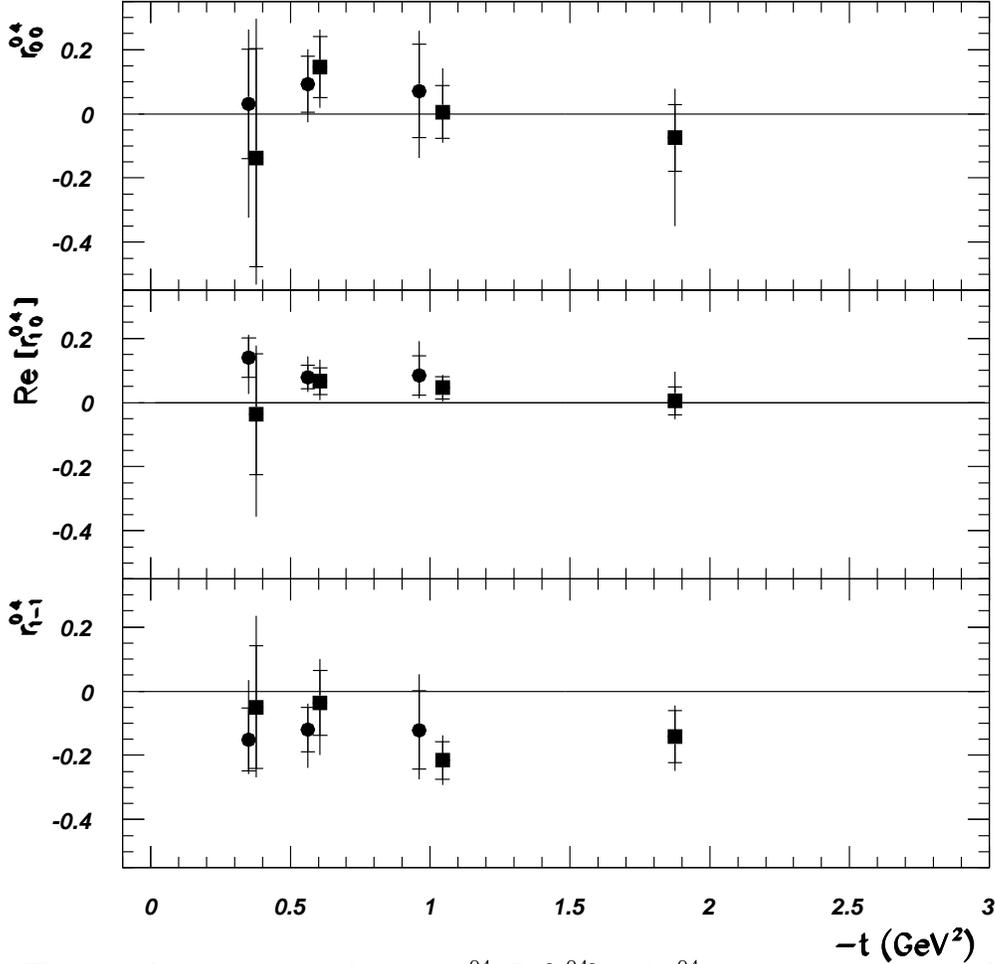,width=0.8\textwidth}
\end{center}
\vspace*{-1.0cm}
\caption{The spin-density matrix elements \rzfzz, 
  Re$[\rzfpz]$ 
and \rzfpm \ 
as a function of $-t$ obtained by fitting Eq.~\ref{hel_2d_dis} 
to the data. The  squares correspond to the 
proton-dissociative sample and the solid 
circles to the elastic sample. 
%the open circles are from earlier ZEUS data 
%\cite{zrho-94}.
The inner error bars indicate the statistical errors, 
the outer bars the statistical and
systematic uncertainties added in quadrature. }
\label{fig:ang_res_2d}
\end{figure}

The three spin-density matrix elements are shown in 
Fig.~\ref{fig:ang_res_2d}, separately for elastic and 
proton-dissociative \rhoz\ production as a function of $-t$.
In the $t$ range of this analysis, 
Re$[\rzfpz]$ tends to be non-zero and positive, 
while \rzfpm\ tends to be non-zero and negative. This is an
indication for small deviations from SCHC giving rise to non-zero
helicity single-flip and double-flip amplitudes.
The element \rzfzz\ is measured to be zero within errors. 

The present results are
not corrected for the non-resonant $\pi^+\pi^-$ production. Hence, strictly
speaking, they only apply to the reaction $\gamma
p(N)\to\pi^+\pi^-p(N)$ 
within the quoted \Mpp\ range. 
However, in the previous section it was shown that the non-resonant
contribution is small and decreases rapidly with $-t$.
In order to assess the sensitivity of the data to changes in the
selected \Mpp\ range, the events with $\Mpp<M_\rho$ and $\Mpp>M_\rho$
were analyzed separately.  No statistically significant effect was
observed. Data at lower $-t$ ($<-t> \approx $0.1 GeV$^2$) from
ZEUS~\cite{zrho-94} show no evidence for a violation of SCHC.
However, lower energy ($<W> \approx 4.3$ GeV) elastic photoproduction
data~\cite{ballam-73}, while showing no SCHC violation for $-t < 0.2$
GeV$^2$, yield positive values for Re$[\rzfpz]$ and negative values
for \rzfpm\ for $-t> 0.2$ GeV$^2$. Recent HERA measurements, both at low
$Q^2$ (0.25--0.85 GeV$^2$)~\cite{zbpc} and higher $Q^2$ (greater than
1 GeV$^2$)~\cite{zbpc,h1schc}, also report similar values, at $<-t>
\approx$ 0.1--0.2 GeV$^2$, although only the low $Q^2$ results are
significantly non-zero.

\subsection{$\phi$ photoproduction}
\subsubsection{Mass distribution}
\begin{figure}[htb]
\vspace*{-0.5cm}
\begin{center}
\epsfig{file=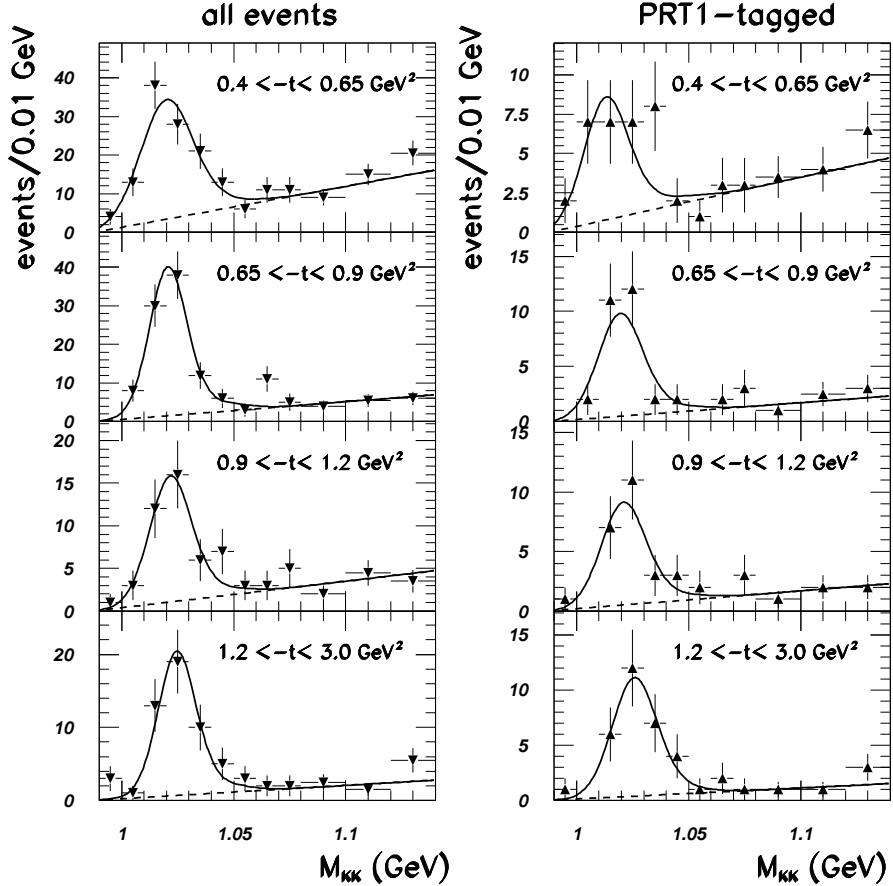,width=0.85\textwidth}
\end{center}
\vspace*{-1.0cm}
\caption{The \Mkk \ distributions in four $-t$ bins. The left-hand plots
are for all events, while the right-hand plots are for PRT1-tagged events.
The points represent the data and the solid 
curves indicate the result of the {f}it discussed in the text. The 
dashed curves represent  the background contribution. Only the 
statistical errors are shown.}
\label{fig:phi_mas_t}
\end{figure}

The invariant mass of the two charged decay products of the $\phi$ is
displayed in Fig.~\ref{fig:phi_mas_t} for different $-t$
ranges for two samples of data: for all events and for those that
have a signal in PRT1 (proton dissociation).
The invariant mass was computed assuming that the two charged
particles were kaons. The lines are fits to a Breit-Wigner 
function, convoluted with a Gaussian resolution function, and a
function describing the background. The background, due mainly to
diffractive $\rho^0$ events for which the two pions are assigned the kaon
mass, was assumed to have the form $\sim (M_{KK} - 2M_K)^\Delta$, where
$\Delta$ is a parameter determined by the fit. For the determination of
the background contribution in each $t$ region, the mass and width
of the $\phi$ were fixed to the PDG values. 
The fit was performed over
a broad mass range $0.99 < M_{KK} < 1.13 $ GeV to give a better
estimate of the background. A narrower mass range $0.99 < M_{KK} < 1.06 $
GeV was then used to select \phiz\ candidates and the background
contribution in this range was subtracted. The mass resolution is about
10~MeV, consistent with the Monte Carlo simulations. 
The background contribution decreases with $-t$, 
from about 23\% at $-t=0.5~ \Gev^2$ to 10\% above $-t=1.2~\Gev^2$.
%\clearpage
\subsubsection{$t-$distributions}
\begin{figure}[htb]
\vspace*{-0.5cm}
\begin{center}
\epsfig{file=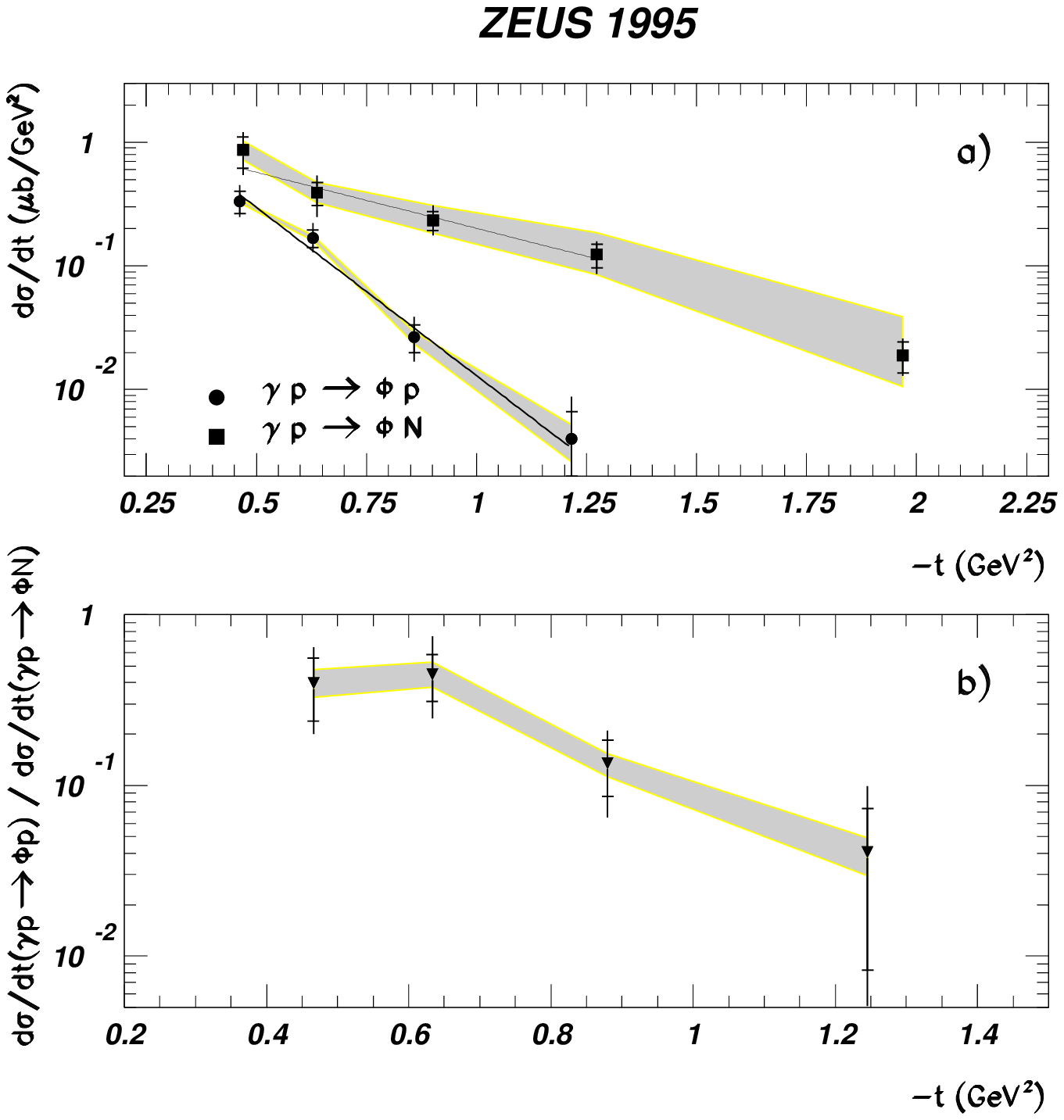,width=0.87\textwidth}
\end{center}
\vspace*{-1.0cm}
\caption{a) The differential cross sections $\dsdt$ for elastic
(circles) and proton-dissociative (squares) \phiz \
photoproduction. The solid lines represent the results of the fit with the
function $Ae^{bt}$. The normalization error of 15\% is not shown.
b) The ratio of the elastic to proton-dissociative cross
sections. The inner error bars indicate the statistical errors,
the outer bars the statistical and systematic uncertainties added 
in quadrature. 
The shaded bands represent the size of the
correlated errors due to the modeling of the proton dissociation in the
Monte Carlo.
}
\label{fig:dsdt-phi}
\end{figure}

The differential cross sections $\dsdt$ for the reactions $\gamma p
\to \phi p$ and $\gamma p \to \phi N$ ($M^2_N<0.1W^2$) were
determined following the separation procedure described in 
Sect.~\ref{sec:sep}; the results are shown in 
Fig.~\ref{fig:dsdt-phi}a. As in
the $\rho^0$ case, both differential cross sections decrease
exponentially with increasing $-t$. A fit with a function of the form
$\dsdt = A\exp(bt)$ in the range $0.4<-t<1.2$~GeV$^2$ yields
$b_{el}$ = 6.3 $\pm
0.7$~(stat.)$\pm 0.6$~(sys.)$\pm 0.3$~(mod.)~GeV$^{-2}$ and
$b_{pd}$ = 2.1 $\pm
0.5$~(stat.)$\pm 0.3$~(sys.)$\pm 0.4$~(mod.) GeV$^{-2}$.  These
values are the same within errors as those obtained for the
$\rho^0$. Note that while $b_{el}^\rho > b_{el}^\phi$ in the region
$-t<$ 0.3 GeV$^2$~\cite{zphi-94}, $b_{el}^\rho \approx b_{el}^\phi$
in the region
0.4 $<-t<$ 1.5 GeV$^2$. There are no data at low $-t$ to make a similar
comparison in the case of the proton-dissociative process,
but for the larger $-t$ range, $b_{pd}^\rho \approx
b_{pd}^\phi$. This indicates that the mass difference between the
$\rho^0$ and the $\phi$ is not important in the large $-t$ \ region.

Fig.~\ref{fig:dsdt-phi}b shows the ratio of the cross sections for
the elastic and the proton-dissociative processes in a given $t$ interval
as function of $-t$. 
The ratio decreases rapidly from unity
at $-t\approx$ 0.5 GeV$^2$ to $\approx 10^{-2}$ at
$-t\approx$ 1.3 GeV$^2$, as in the $\rho^0$ case. 
A fit to the
ratio gives $\Delta b=3.0 \pm
0.8$ (stat.)~$^{+0.5}_{-0.9}$~(sys.)~$^{+0.3}_{-0.5}$~(mod.)~GeV$^{-2}$.
This value is the same within errors as that obtained for the
$\rho^0$, 
and is thus independent of the
type of the vector meson produced at the photon vertex (see
Sect.~\ref{sec:ff}).

\begin{figure}[htb]
\begin{center}
\epsfig{file=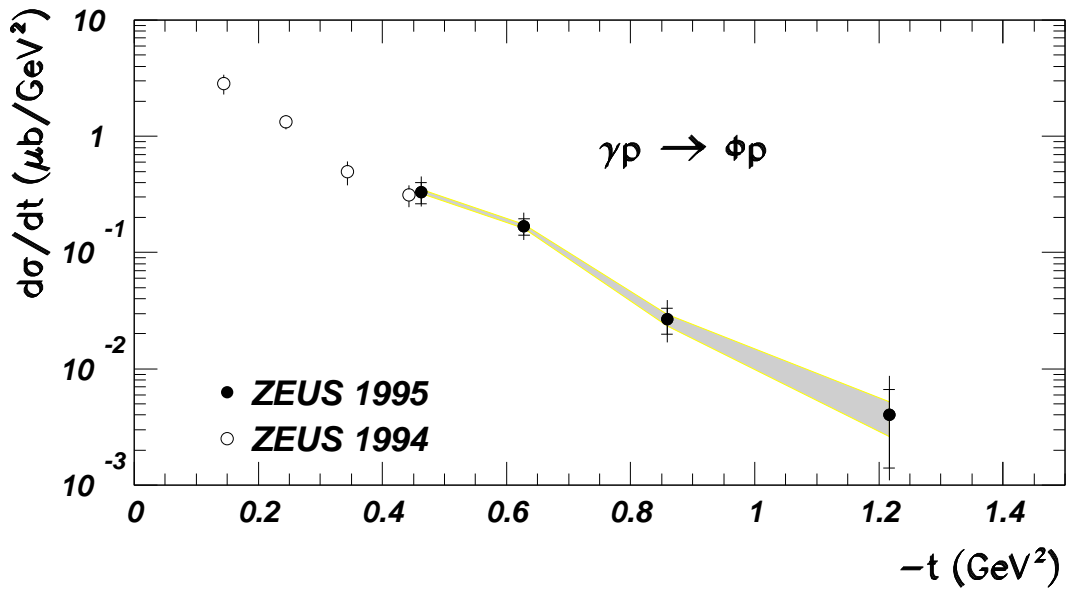,width=\textwidth}
\end{center}
\caption{The differential cross sections $\dsdt $ for elastic
\phiz \ photoproduction. The solid circles correspond to the data of 
this analysis  and
the open circles are from an earlier ZEUS result~\cite{zphi-94}. 
The inner error bars indicate the statistical errors,
the outer bars the statistical and systematic uncertainties
 added in quadrature.
The shaded band represents the size of the correlated errors due
to the modeling of the proton dissociation in the 
Monte Carlo. The normalization error of 15\% is not shown. }
\label{fig:dsdt-phi-el}
\end{figure}

In Fig.~\ref{fig:dsdt-phi-el} the results for $\dsdt$ for
elastic $\phi$ photoproduction are displayed, together with ZEUS
measurements at small $-t$~\cite{zphi-94}. The low $-t$ data have
been rescaled to the present $W$ value by assuming the Pomeron trajectory
as measured in this analysis (see Sect.~\ref{trajectory}); 
both data sets are plotted
with statistical errors only. The results are consistent.

\clearpage
\subsection{$J/\psi$ photoproduction}
\subsubsection{Mass distribution}
\begin{figure}[htb]
\vspace*{-1.0cm}
\begin{center}
\epsfig{file=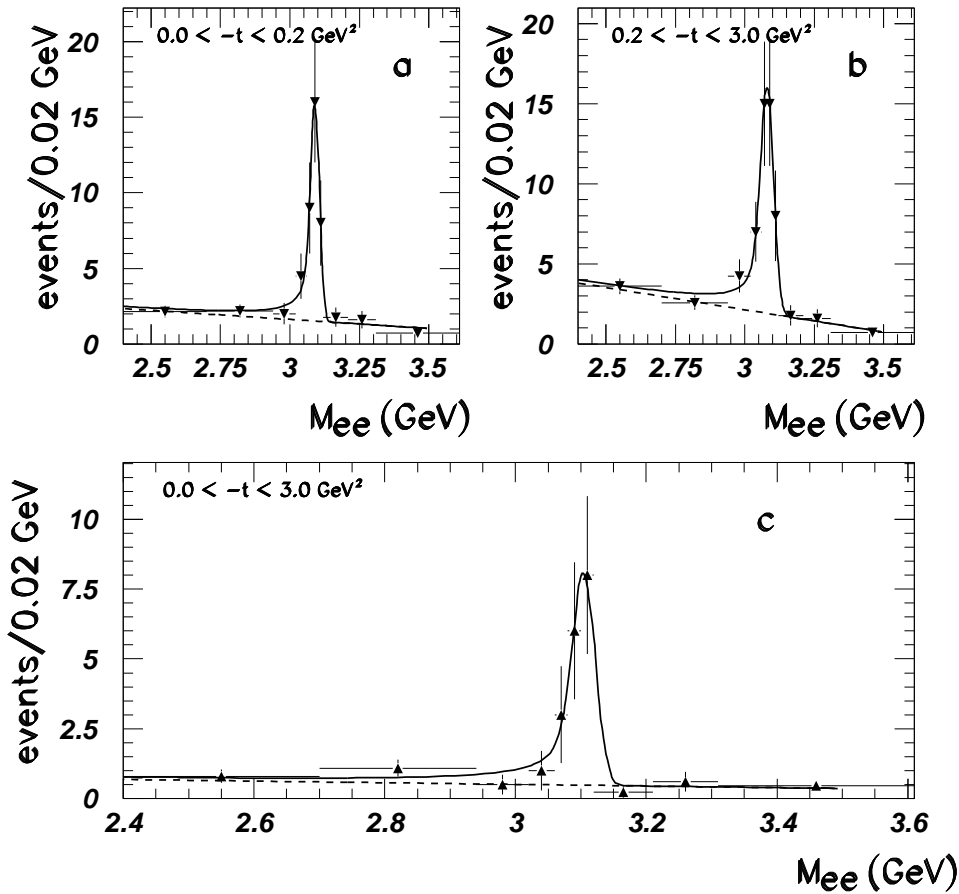,width=0.85\textwidth}
\end{center}
\vspace*{-1.2cm}
\caption{The \Mee \ distributions in two $-t$ bins for the whole
sample (a,b) and in one bin for the PRT1 tagged events (c). 
The points represents the data and the curves indicate 
the result of the fit discussed in the text. The dashed curve represents  the
background contribution. }
\label{fig:jpsi-mass}
\end{figure}

Figure~\ref{fig:jpsi-mass} shows the two body
invariant mass distribution in the region 2.4--3.6 GeV, for the whole
sample in two $-t$ ranges and for the PRT1-tagged sample in one $-t$ bin. 
No lepton identification was performed for this sample. 
Although the events therefore
represent a sum of $e^+e^-$ and $\mu^+\mu^-$ final
states
the electron mass was always assumed for each of the two charged 
particles\footnote{Assuming the muon mass
does not change any of the results.}. 
A narrow peak is observed around 3.1 GeV. Fits were made
to the invariant mass distribution using the sum of a Gaussian 
(for the $\mu$ case), a Gaussian (modified by the energy loss spectra 
due to bremsstrahlung in dead material before the CTD for the $e$ case), 
and a background which linearly decreases with mass. Equal numbers of $e$ 
pairs and $\mu$ pairs were assumed.
The fitted $J/\psi$ mass is in good agreement with the PDG value. 
The background in the $J/\psi$ mass
region (2.87--3.13 GeV) is (20 $\pm$ 5)\%, independent of
$t$ in the range under study ($-t<$ 3 GeV$^2$).
$\ $ 
\vspace*{0.1cm}
$\ $

\subsubsection{$t-$distributions}
\begin{figure}[htb]
\vspace*{-0.5cm}
\begin{center}
\epsfig{file=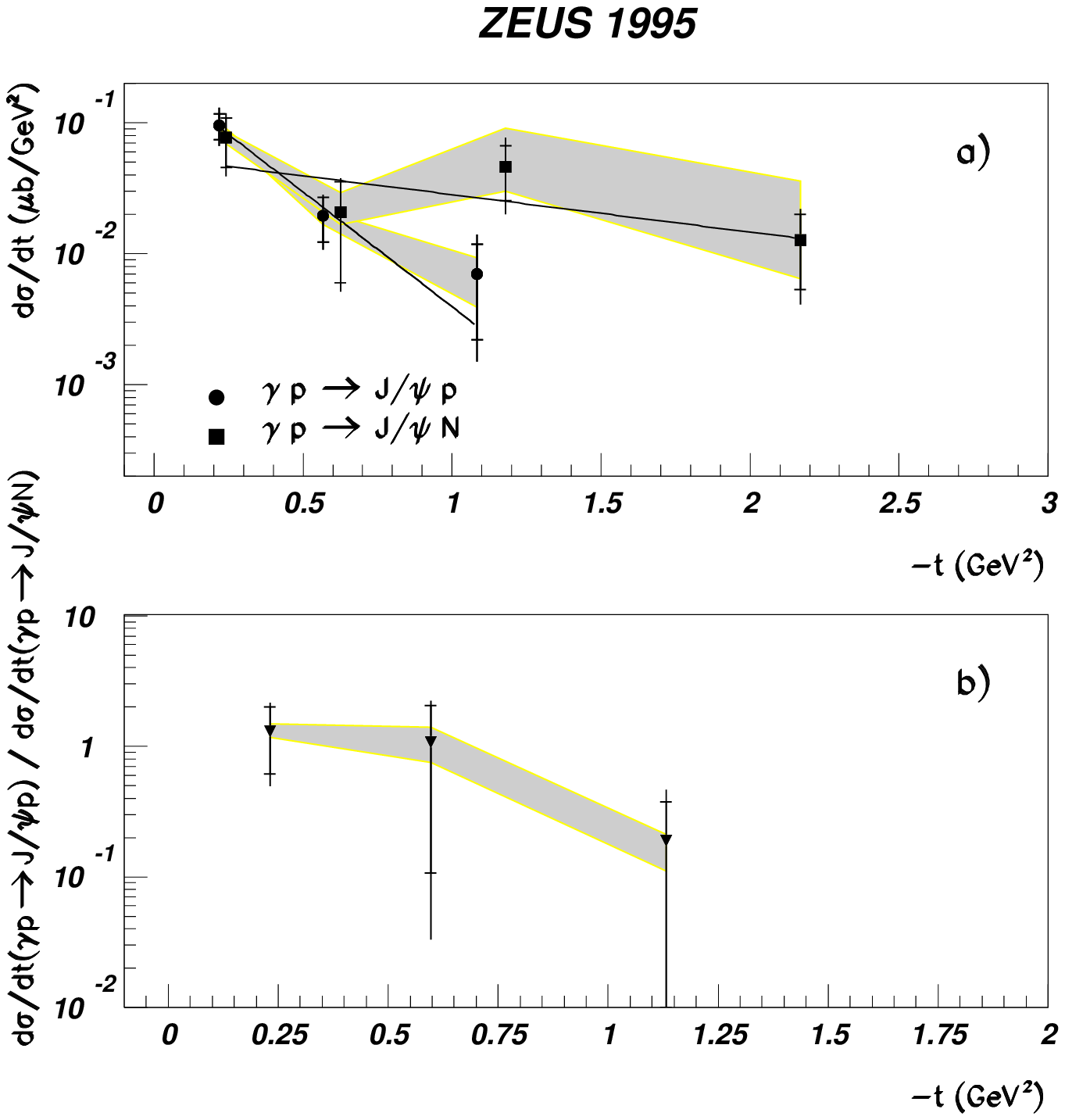,width=0.8\textwidth}
\end{center}
\vspace*{-1.0cm}
\caption{a) The differential cross sections $\dsdt $ for elastic
(circles) and proton-dissociative (squares) \jpsi \
photoproduction. 
The solid lines represent the results of the fit with the function
$Ae^{bt}$.  The normalization error of
15\% is not shown.
b) The ratio of the elastic to the proton-dissociative cross
sections. The inner error bars indicate the statistical errors,
the outer bars the statistical and systematic uncertainties 
 added in quadrature.
The shaded bands represent correlated errors due
to the modeling of the proton dissociation in the Monte Carlo.
}
\label{fig:dsdt-jpsi}
\end{figure}

The differential cross sections $\dsdt$ for the elastic and the
proton-dissociative $J/\psi$ photoproduction reactions are shown in
Fig.~\ref{fig:dsdt-jpsi}a.  
In contrast to the $\rho^0$ and $\phi$ cases, the high mass of the
$J/\psi$  and the resulting large opening angle of the decay particles
result in significant acceptance in the low $-t$ region. 
However, in the case of the proton-dissociative process, the
effect of the minimum kinematically allowed $-t$, $-t_0\approx
0.06~{\rm GeV^2}$ (see Eq.~\ref{Eq:ptrel}), prevents
the measurement of \dsdt\ in the low $-t$ bins.  The elastic
differential cross section falls exponentially, with 
$b_{el}^{J/\psi}$ = 4.0 $\pm
1.2$(stat.)$^{+0.7}_{-1.1}$(sys.)$^{+0.4}_{-0.6}$(mod.)~GeV$^{-2}$, while
for the proton-dissociative reaction $b_{pd}^{J/\psi}$
= 0.7 $\pm 0.4$(stat.)$\pm 0.2$~(sys.)$^{+0.5}_{-0.3}$(mod.)~GeV$^{-2}$.
The slope 
$\Delta b$ agrees, within errors, with the values for the 
$\rho^0$ and $\phi$.

Fig.~\ref{fig:dsdt-jpsi}b shows the ratio of the elastic
to the proton-dissociative differential cross sections. This ratio also
falls from a value of about 1 at low $-t$ to a value of about 0.1
for $-t>$ 1 GeV$^2$, similar to the $\rho^0$ and the $\phi$
cases.

\begin{figure}[htb]
\vspace*{-0.3cm}
\begin{center}
\epsfig{file=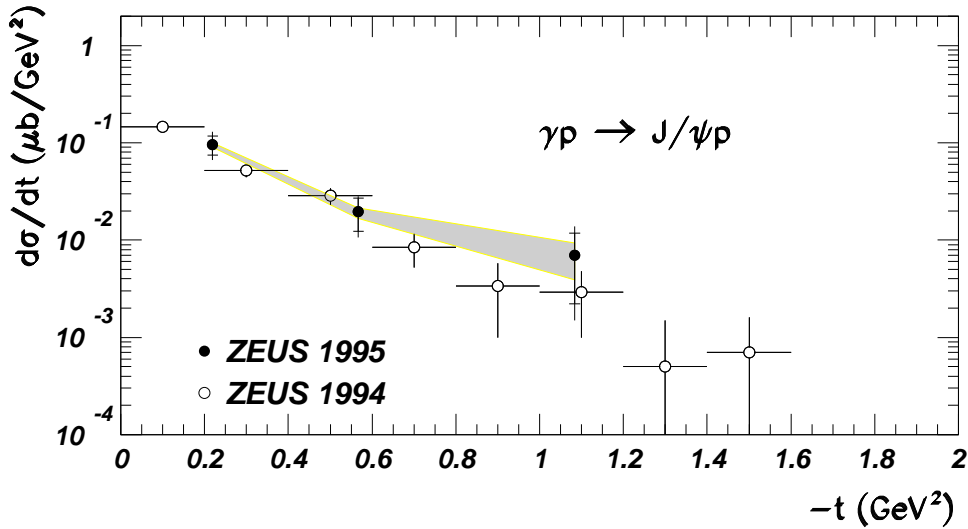,width=0.9\textwidth}
\end{center}
\vspace*{-.2cm}
\caption{The differential cross section $\dsdt $ for elastic
$J/\psi$ \ photoproduction. The solid circles correspond to the present
measurement, the open circles to the published ZEUS untagged
photoproduction results
\cite{zjpsi-94}. 
The shaded band represents correlated errors due to the modeling of the 
proton dissociation in the Monte Carlo. The normalization error of
15\% is not shown. 
}
\label{fig:dsdt-jpsi-zeus-h1}
\end{figure}

The present measurements of $\dsdt$ for the reaction $\gamma p \to
J/\psi p$ are shown in Fig.~\ref{fig:dsdt-jpsi-zeus-h1}, together with
our earlier untagged (i.e.\ $Q^2 \sim 0$) 
photoproduction data~\cite{zjpsi-94} ($\langle
W\rangle = 90$ GeV). The data sets agree within errors.

\section{Comparisons of data to models}

\subsection{Cross section comparisons with the pQCD-based models}
\begin{figure}[htb]
\vspace{-0.3cm}
\begin{center}
\epsfig{file=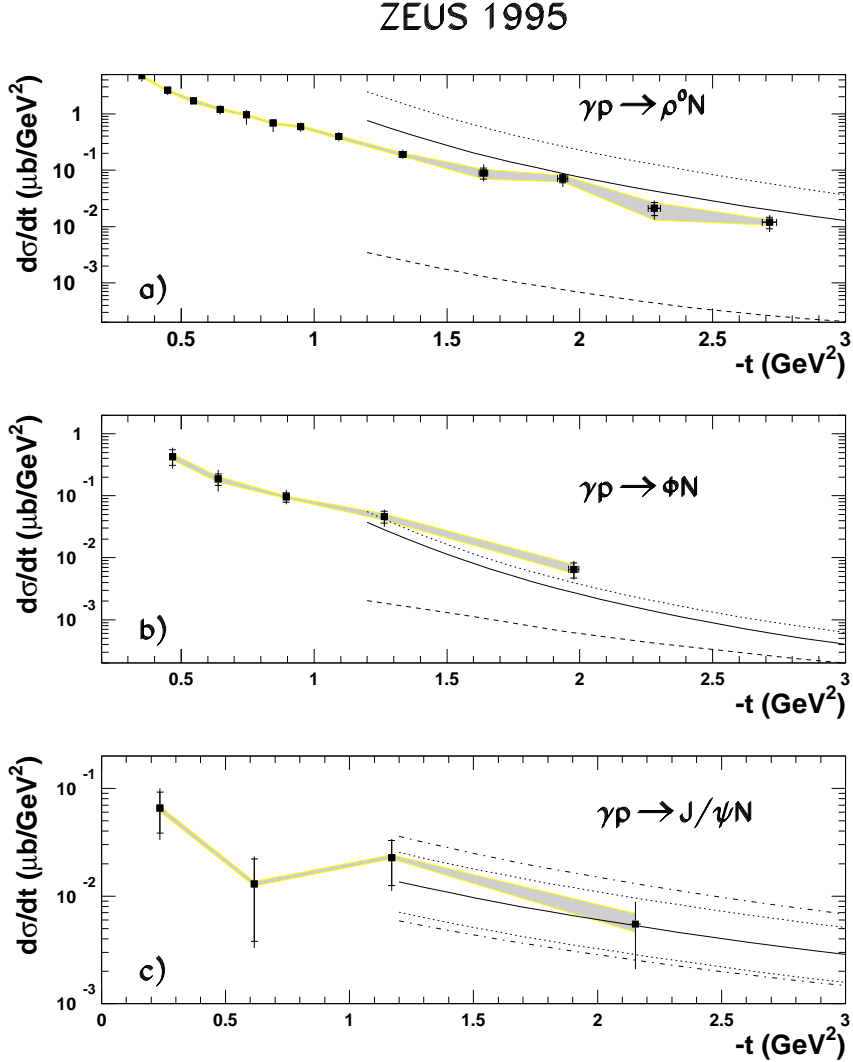,width=0.75\textwidth}
\end{center}
\vspace{-1.2cm}
\caption{Comparison of the measured differential cross sections
\dsdt for proton-dissociative vector-meson production for
$M_N^2<0.01W^2$ with the pQCD-based models for a)~\rhoz\ , 
b)~\phiz\ ~\cite{ivanov} and c) $J/\psi$ ~\cite{bartels}.
The inner error bars indicate the statistical errors,
the outer bars the statistical and systematic uncertainties 
added in quadrature. The shaded bands represent correlated errors due
to the modeling of the proton dissociation in the Monte Carlo. The 
normalization error of 15\% is not shown. For the description of the 
curves see text. }
\label{fig:pqcd}
\end{figure}

In order to compare the data to pQCD predictions for light
\cite{ginzburg} and heavy \cite{bartels} mesons, the cross sections
were redetermined in the region of validity of the model calculation,
{\it viz.} $M_N^2<0.01W^2$, using the EPSOFT MC simulation. The
measured proton-dissociative cross sections for the three vector
mesons are shown in Fig.~\ref{fig:pqcd}.

The calculations \cite{ginzburg} for the production of the 
\rhoz\ and \phiz\ mesons
were performed at lowest order in $\alpha_S$.  Only helicity non-flip
and single-flip amplitudes were taken into account.
The non-perturbative effects 
were simulated by introducing
effective quark masses. In Fig.~\ref{fig:pqcd}a and \ref{fig:pqcd}b, the solid
(dotted) curves represent results for a quark mass of 300 (200)~MeV. 
For the calculation of the \rhoz\ cross-section,  the
sub-asymptotic  \rhoz\ wave-function was used, whereas for
the \phiz\ production the asymptotic one
\cite{ginzburg,ivanov} was employed. 
The contribution of the perturbative cross section,
represented by the dashed lines in Fig.~\ref{fig:pqcd}a and \ref{fig:pqcd}b,  
is well below the \rhoz\ and $\phi$ data.
This observation, together with the helicity analysis (see Sect.~\ref{sec:hel})
in which no significant production of \rhoz\ mesons with helicity 0
is observed, implies that these perturbative calculations are not 
applicable in this regime.

The situation is different in the $J/\psi$ case.  The perturbative
QCD prediction in the LLA \cite{bartels} expansion in terms of 
$\ln{(W^2/W_0^2)}$ (equivalent to $\ln{1/x}$)
compares satisfactorily with the data (solid line in Fig.~\ref{fig:pqcd}c).
A value of $\alpha_S$=0.2 and an energy scale $W_0=1$~GeV 
have been used in this calculation. 
It should be
noted, however, that the 
uncertainties due to the choice of $\alpha_S$ ($\pm10\%$ -- dotted
lines) and the $W_0$ scale ($0.2<W_0^2<5$~GeV$^2$ -- dashed-dotted
lines) are significant. 

\subsection{Ratios of cross sections for vector-meson photoproduction}
\label{sec:ratios}
Flavor independence predicts that the ratio of the
production cross sections of $\phi$:$\rho^0$ should be 2:9 and that of
$J/\psi$:$\rho^0$ should be 8:9. These predictions are in striking
disagreement with the previously published low-$Q^2$
data: at $W$=70 GeV, the ratios 
%\begin{eqnarray*}
$\phi$:$\rho^0=0.065 \pm 0.013$~\cite{zphi/rho} 
%\end{eqnarray*}
and \linebreak
%\begin{eqnarray*}
$J/\psi$:$\rho^0=0.00294\pm0.00074$~\cite{zjpsi-94}
%\end{eqnarray*} 
were measured. 
In contrast, in the DIS
kinematic region, for $Q^2\approx$ 12 GeV$^2$, 
the ratios 
$\phi$:$\rho^0$=0.18$\pm$0.05~\cite{zphi/rho} 
and 
0.19$\pm$0.04~\cite{h1phi/rho} 
were obtained at $W \approx$ 100 GeV.  This analysis gives
the ratio $J/\psi$:$\rho^0 0.64\pm 0.35$ for $Q^2$ = 10 GeV$^2$
and 1.3$\pm$0.5 for $Q^2$ = 20 GeV$^2$~\cite{h1jpsi/rho} at $W
\approx$ 100 GeV. These results suggest that flavor independence
may hold at large $Q^2$.

\begin{figure}[htb]
\vspace{-0.5cm}
\begin{center}
\epsfig{file=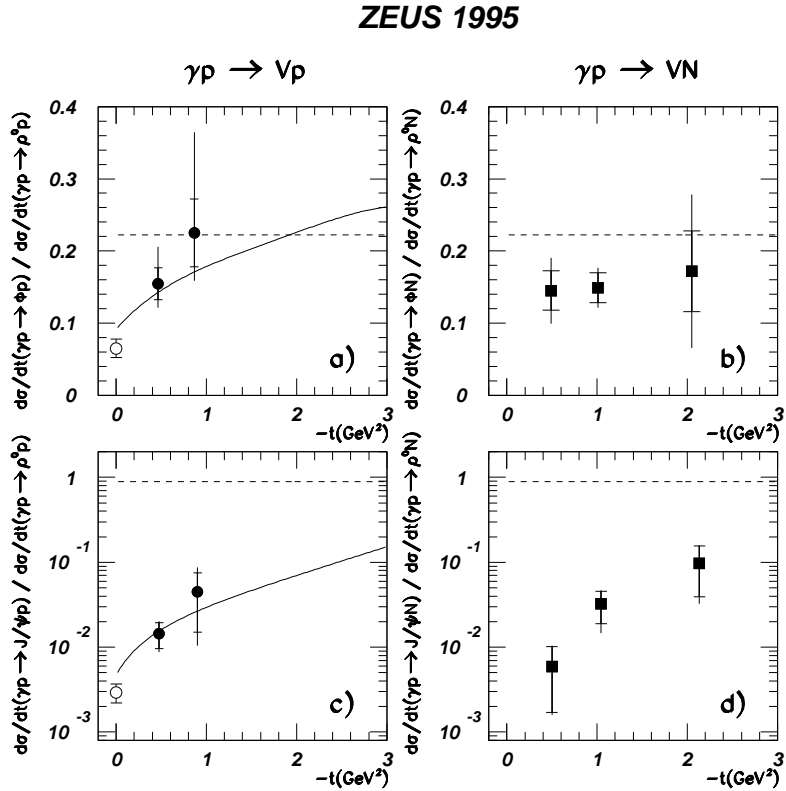,width=0.73\textwidth}
\end{center}
\vspace{-1.0cm}
\caption{The ratios of the cross sections \dsdt\ for 
 \phiz \ to \rhoz \ (a and c) and \jpsi \ to \rhoz \ (b
and d)  
for elastic and proton-dissociative
 photoproduction. 
The inner error bars indicate the statistical errors,
the outer bars the statistical and systematic uncertainties
 added in quadrature.
The ratios at $t=0$ were obtained from the
data\cite{zphi/rho,zjpsi-94} as 
explained in the text.
The solid curves represent
predictions of a non-perturbative QCD model \cite{ref:isabelle} for
elastic 
photoproduction. 
The dashed lines correspond to the expectations of flavor independence.}
\label{fig:rVtorho}
\end{figure}

The ratios $\phi:\rhoz$ and $J/\psi:\rhoz$ of the differential cross sections 
for elastic reactions from this analysis
are plotted in Fig.~\ref{fig:rVtorho}a and \ref{fig:rVtorho}c  
as a function of $-t$.
For both ratios, the point  at $t=0$ was obtained by using 
the ratios of values of the total 
elastic cross section and rescaling them by the ratios
of the corresponding slopes, $b$, of the differential cross sections,
using the relation $\dsdt(t=0)=b\sigma$. An increase of the $\phi:\rhoz$ 
ratio up to $-t \approx$ 1 GeV$^2$ is observed, approaching the 
expected 2:9 ratio.   Although the
$J/\psi:\rhoz$ ratio increases quickly up
to $-t \approx $ 0.5 GeV$^2$, it remains more than an order of 
magnitude smaller than the 8:9 expectation. A non-perturbative QCD 
model \cite{ref:isabelle}
successfully describes the ratios for the elastic production at 
$-t>$ 0.3 GeV$^2$.
 
The ratios for the proton-dissociative reactions are shown in
Fig.~\ref{fig:rVtorho}b and \ref{fig:rVtorho}d. 
There are no data at $t$ = 0 for this
process. Both ratios are consistent with the corresponding elastic results.
At $-t\approx 2.2$~GeV$^2$, the  $J/\psi:\rhoz$ ratio is still well 
below the 8:9 expectation.

\subsection{Test of Regge factorization hypothesis}
\label{sec:ff}

The hypothesis of Regge
factorization~\cite{gribov-fact} implies that the ratio of the 
elastic to proton-dissociative differential cross sections should
be independent of the type of vector meson produced at the photon
vertex. Furthermore, this ratio should be the same as for 
hadron-proton elastic and proton-dissociative processes.

Several results discussed in previous sections are consistent with
this hypothesis: the decrease in the $\pi\pi$ resonance-shape
distortion with increasing $-t$ is the same for the elastic and
proton-dissociation channels; the $\rho^0$ spin-density matrix
elements for both channels agree; and the difference ($\Delta
b$) between the slopes of the $t$-distributions for elastic and
proton-dissociation processes is independent of the type of vector
meson ($\rho^0, \phi, J/\psi$) produced at the photon vertex.

\begin{figure}[htb]
\vspace{-0.5cm}
\begin{center}
\epsfig{file=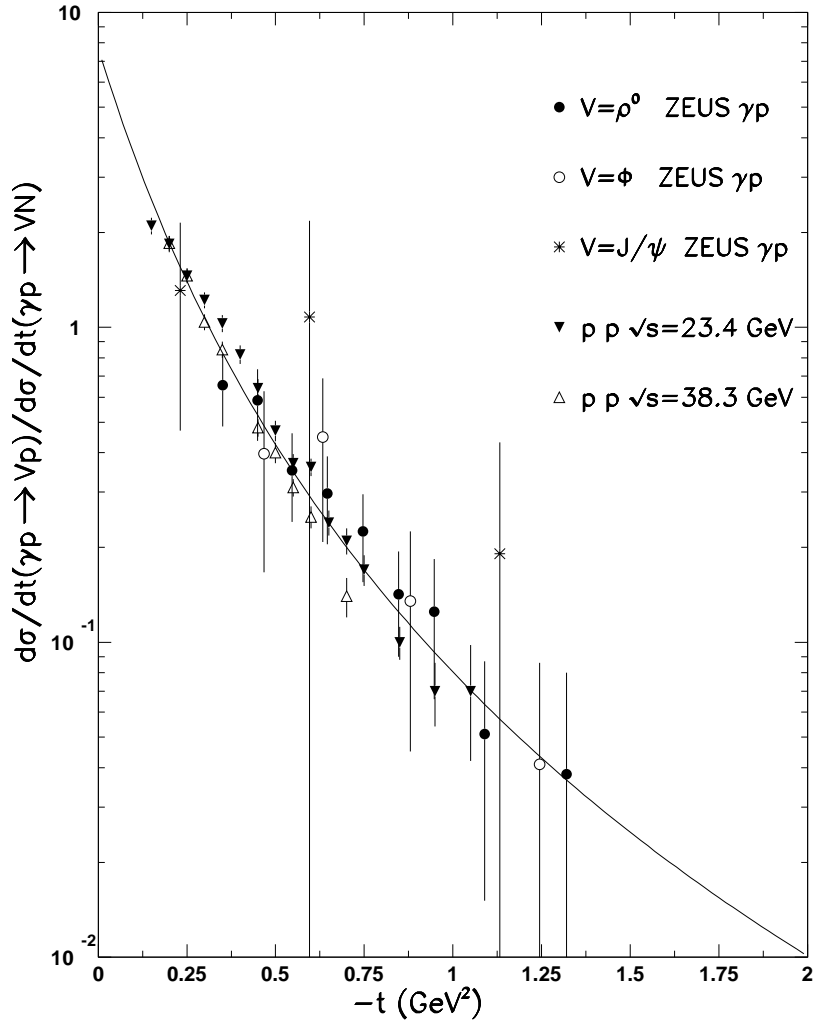,width=0.71\textwidth}
\end{center}
\vspace{-1.0cm}
\caption{The ratio of the elastic to proton-dissociative differential 
cross sections as a function of $-t$ for vector-meson photoproduction,
together with data from $p p$ reactions~\cite{chlm} at
$\sqrt{s}$ = 23.4 and 38.3 GeV. The curve is the result of a combined fit 
to all the data as explained in the text.}
\label{fig:ff}
\end{figure}

To test this hypothesis further, the ratios
$\frac{d\sigma}{dt}(\gamma p \to V p)/\frac{d\sigma}{dt}(\gamma p \to
V N)$ for the three vector mesons $V = \rho^0, \phi, J/\psi$
are shown in Fig.~\ref{fig:ff}. As can
be seen, the ratio agrees within errors for each of the three
vector mesons. In the same
figure, the ratio $\frac{d\sigma}{dt}(p p \to p
p)/\frac{d\sigma}{dt}(p p \to p N)$ at center of mass energies of
$\sqrt{s}$ = 23.4 and 38.3 GeV~\cite{chlm} is also shown. The $p p$
proton-dissociative reaction is defined for $M_N^2 \le 0.05 s$. The
ratios for the $p p$ reactions are in agreement with those of
the vector-meson photoproduction. These observations 
confirm the factorization hypothesis.

In a naive additive quark
model~\cite{kokkedee}, the ratio shown in Fig.~\ref{fig:ff} measures
the proton form factor. In such a model, the differential cross
sections can be related to the form factor of the hadrons involved in
the scattering process. In the framework of the VDM, this
relation can also be used for the vector-meson photoproduction reactions 
discussed above. The elastic and proton-dissociative processes can be
expressed as
\begin{eqnarray}
\left(\frac{d\sigma}{dt}\right)_{el}
\sim F_p^2\ F_V^2\ |\cal{A}|^{\rm 2}, \nonumber \\
\left(\frac{d\sigma}{dt}\right)_{pd} \sim \ F_V^2\ |\cal{A}|^{\rm 2},
\end{eqnarray}
where $\cal{A}$ is the amplitude describing the constituent
interaction, the form factor of the proton is indicated by $F_p$ and that
of the vector meson by $F_V$. Thus, the ratio of the elastic and the
proton-dissociative cross sections gives the proton form factor:
\begin{equation}
\frac{\frac{d\sigma}{dt}(\gamma p \to V p)}
{\frac{d\sigma}{dt}(\gamma p \to V N)} = F_p^2 =
\frac{D}{(1-t/m^2)^4},
\label{eq:ff}
\end{equation}
where the electromagnetic dipole expression for the
proton form factor has been used and $D$ and $m$ are free parameters.

Equation~\ref{eq:ff} was fitted to both the
photoproduction and $pp$ data and the result is shown 
as the curve in Fig.~\ref{fig:ff}.
The best
fit yielded the parameters $D$ = 7.7 $\pm$ 0.4 and $m^2$ = 0.47
$\pm$ 0.01 GeV$^2$, and gives a fair description of the data. 
Note that the
mass scale parameter obtained by this fit is smaller than the value of
0.71 GeV$^2$ measured from electron--proton elastic scattering..
Fitting only to the ZEUS photoproduction data yields $D=11.2\pm8.0 $ and
$m^2=0.43\pm0.13~\Gev^2$.
\subsection{Determination of the Pomeron trajectory}
\label{trajectory}
The measurement of the variation of the energy dependence of the
elastic cross section with momentum transfer $t$ yields a direct
determination of the Pomeron trajectory, as shown in
Eq.~\ref{eq:regge}. 
Such an analysis is presented for elastic \rhoz\ and \phiz\ photoproduction.

Since the reaction $\gamma p \to \rho^0 p$ is dominated by Pomeron
exchange only at high energies, \dsdt\ measurements at very low $W$ cannot 
be used in this analysis. Therefore, the only
fixed target experiment that can be used is the OMEGA experiment
(WA4)~\cite{omega}, which measured elastic photoproduction at $W$ =
 8.2 and 10.1 GeV in the $t$ range $0.06<-t<1$ GeV$^2$.
The measurement of the H1 collaboration~\cite{h1rho-94} at $W$=55
GeV, the earlier measurements of the ZEUS collaboration at $W$=71.2
GeV~\cite{zrho-94} and at $W$=73 GeV~\cite{zrho-lps}, and the present
data at $W$=94 GeV are also used.

\begin{figure}[htb]
\begin{center}
\epsfig{file=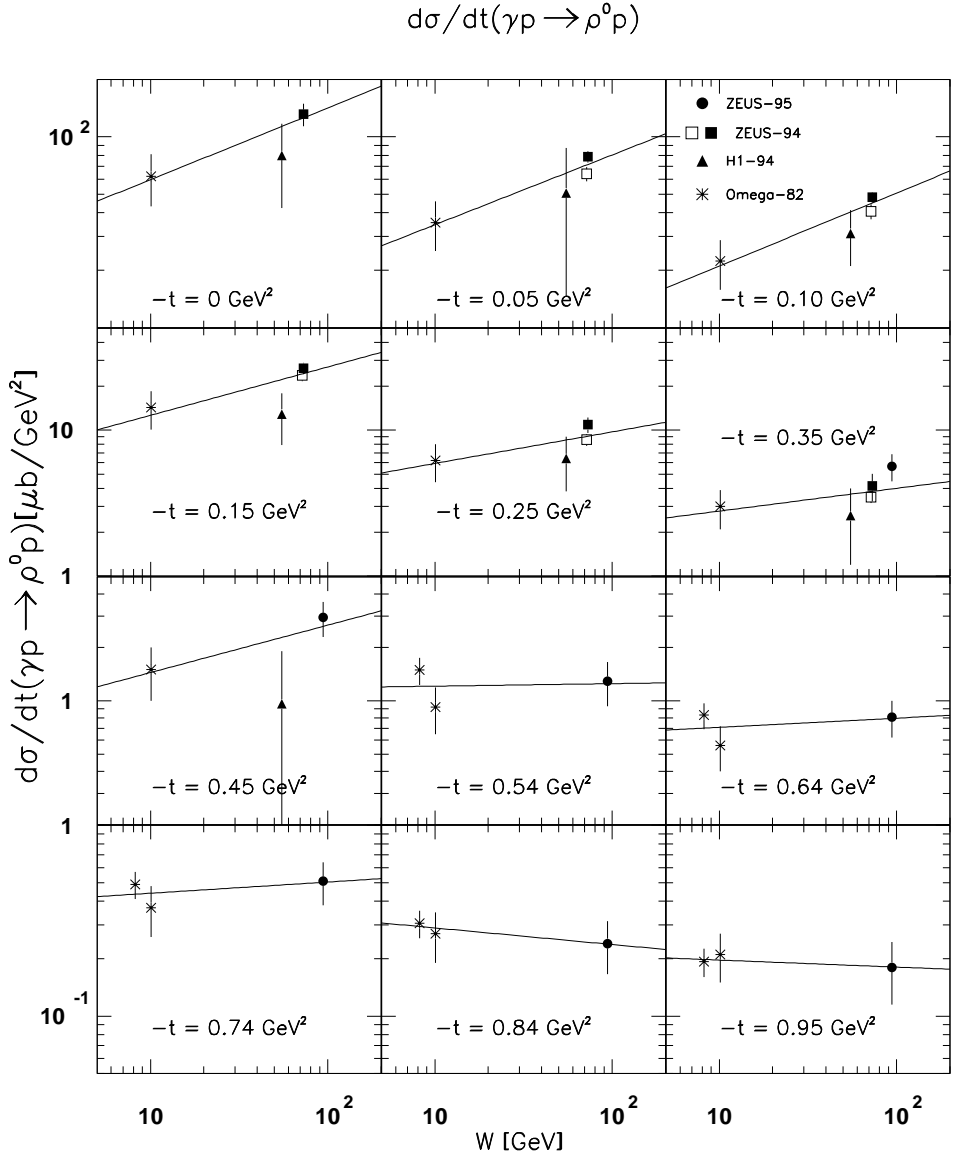,width=0.84\textwidth}
\end{center}
\caption{Cross sections for exclusive \rhoz\ production from ZEUS, H1 and 
OMEGA~\cite{zrho-94,zrho-lps,h1rho-94,omega} at fixed $-t$ 
values as a function of $W$. 
The error bars show the statistical and systematic uncertainties added in 
quadrature. The lines correspond to the results of the fits to 
d$\sigma/$d$t\propto (W^2)^{2\alphapom(t)-2}$. 
}
\label{fig:dsdtw-rho}
\end{figure}

\begin{figure}[htb]
\begin{center}
\epsfig{file=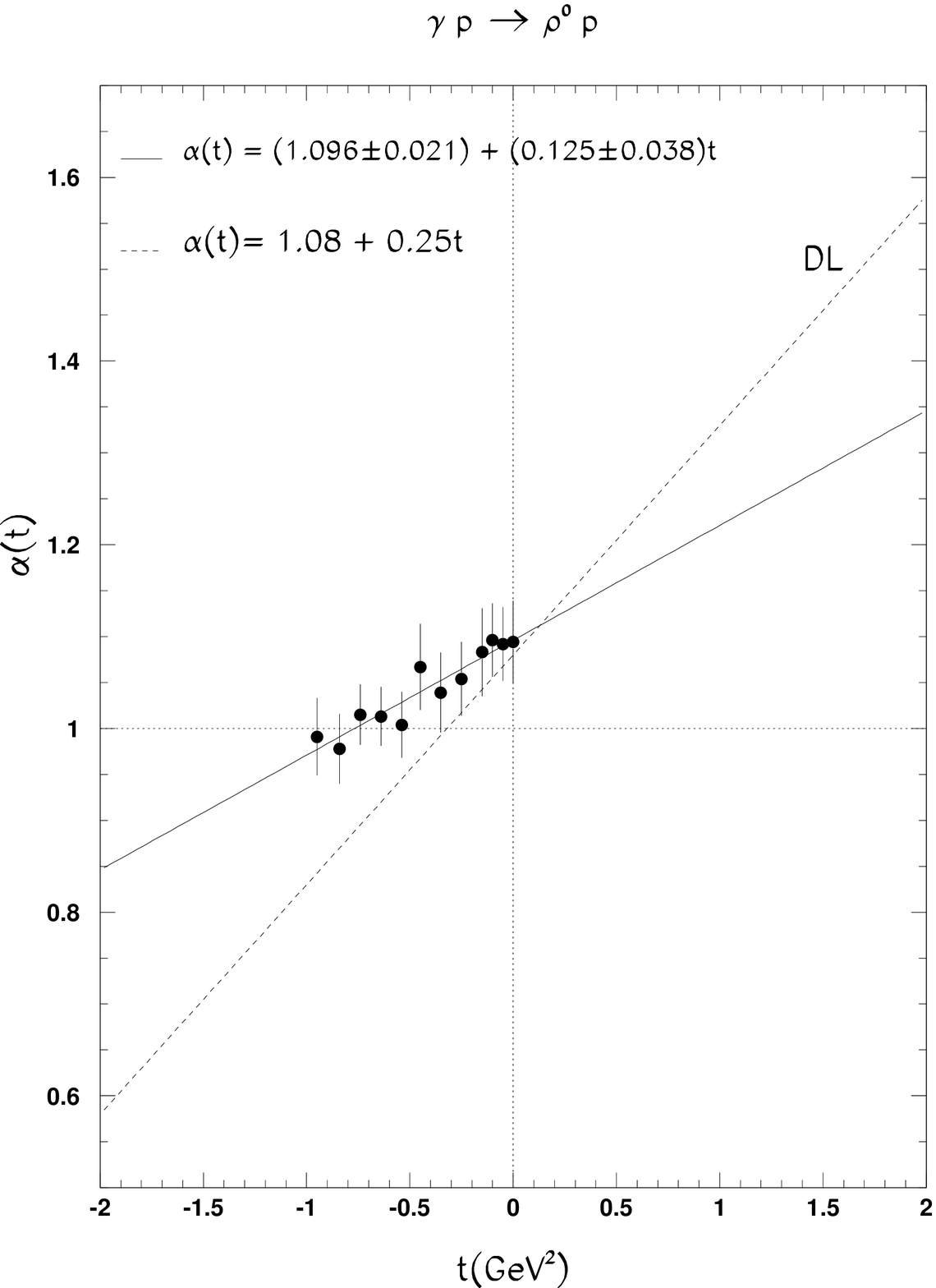,width=0.9\textwidth}
\end{center}
\caption{ 
Determination of the Pomeron trajectory from the reaction $\gamma p \to
\rho^0 p$. The dots are the values of the trajectory at a given $-t$ as
determined from Fig.~\ref{fig:dsdtw-rho} and the solid line is the result
of a linear fit to these values. 
The Pomeron trajectory as determined by
DL~\cite{dl} is shown for comparison as a dashed line. 
}
\label{fig:alpha-rho}
\end{figure}

The $\dsdt$ data used in the determination of $\alphapom(t)$ are
presented in Fig.~\ref{fig:dsdtw-rho} in 12 $t$ bins in the range $0
\le -t \le 0.95$ GeV$^2$. The errors are the statistical and
systematic uncertainties combined in quadrature. The line in each $t$ bin
is the result of a fit of the form $W^n$, where $n = 4\alphapom(t) - 4$.
The resulting values of $\alphapom(t)$ are plotted in
Fig.~\ref{fig:alpha-rho} as a function of $t$. A linear fit to the
data yields
\begin{equation}
\alphapom(t) = (1.096 \pm 0.021) + (0.125 \pm 0.038) t,
\label{eq:rho_trajectory}
\end{equation}
and is plotted as a full line in the figure. The quality of these fits
is acceptable.
The dashed line is the
Pomeron trajectory 1.0808 + 0.25 $t$, as determined by Donnachie and
Landshoff (DL)~\cite{dl}. The resulting intercept, $\alphapom(0)$, is in
excellent agreement with that of DL; however, the slope, $\alphappom$,
is smaller in the present determination. 

\vspace{0.5cm}

\clearpage
The elastic photoproduction of $\phi$ mesons is a good reaction with 
which to study the
properties of the Pomeron, since this is the only trajectory that can be
exchanged~\cite{fruend}, assuming 
the $\phi$ to be a pure $s\bar{s}$ state. This
allows the use of data at very low $W$.

\begin{figure}[htb]
\vspace{-0.3cm}
\begin{center}
\epsfig{file=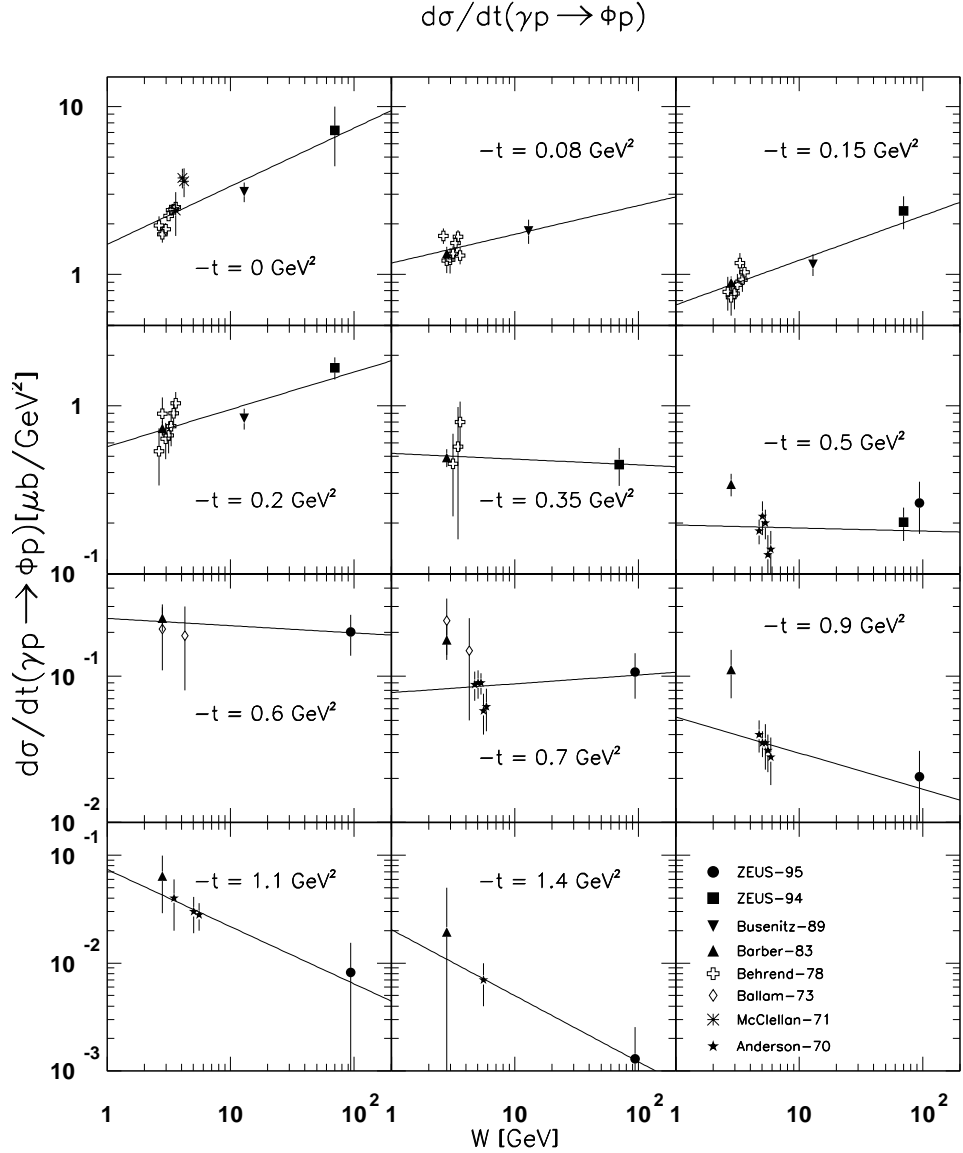,width=0.84\textwidth}
\end{center}
\vspace{-0.6cm}
\caption{Cross sections for exclusive \phiz\ production from ZEUS and 
low energy measurements~\cite{zphi-94,behrend-78,barber-83,ballam-73,
mcclellan-71,anderson-70,busenitz-89} at fixed $-t$ 
values as a function of $W$. 
The error bars show the statistical and systematic uncertainties added in 
quadrature. The lines correspond to the results of the fits to 
d$\sigma/$d$t\propto (W^2)^{2\alphapom(t)-2}$. 
}
\label{fig:dsdtw-phi}
\end{figure}

The $\dsdt$ data used for the trajectory determination include
the following: $W$=2.64--3.60 GeV~\cite{behrend-78},
$W$=2.8~\cite{barber-83}, $W$=2.81--4.28 GeV~\cite{ballam-73},
$W$=3.59--4.21 GeV~\cite{mcclellan-71}, $W$=4.73--5.85
GeV~\cite{anderson-70}, $W$=12.89 GeV~\cite{busenitz-89}, $W$=70
GeV~\cite{zphi-94} and the present measurement at $W$=94 GeV. These
data points are displayed in Fig.~\ref{fig:dsdtw-phi} for 11 $t$
values in the range $0 \le -t \le 1.4$ GeV$^2$. 
The lines are the
results of fits of the function $W^n$ to the data.
\clearpage

\begin{figure}[htb]
\vspace{-0.5cm}
\begin{center}
\epsfig{file=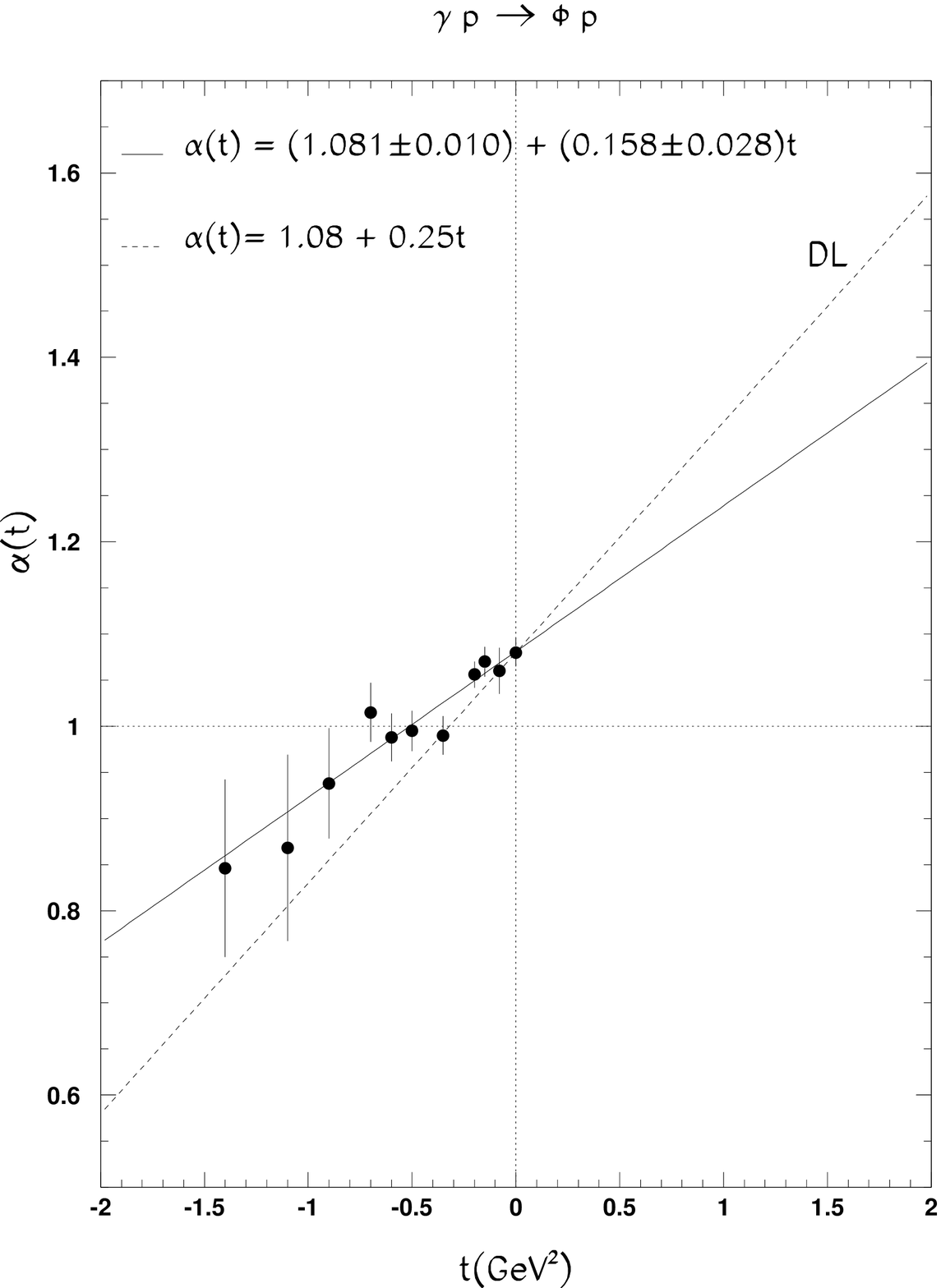,width=0.78\textwidth}
\end{center}
\vspace{-0.5cm}
\caption{ 
Determination of the Pomeron trajectory from the reaction $\gamma p \to
\phi p$. The dots are the values of the trajectory at a given $-t$ as
determined from Fig.~\ref{fig:dsdtw-phi} and the full line is the result
of a linear fit to these values.
The Pomeron trajectory as determined by
DL~\cite{dl} is shown for comparison as a dashed line. 
}
\label{fig:alpha-phi}
\end{figure}

The resulting values of the trajectory $\alphapom(t)$ are shown in
Fig.~\ref{fig:alpha-phi} as a function of $t$. Assuming a linear
trajectory,
\begin{equation}
\alphapom(t) = (1.081 \pm 0.010) + (0.158 \pm 0.028) t
\label{eq:phi_trajectory}
\end{equation}
is obtained. Again the quality of the fits is generally acceptable.
This trajectory is shown as a solid line in the figure and compared
with the DL trajectory, which is plotted as a dashed line. 
\clearpage

As can be seen from (\ref{eq:rho_trajectory}) and (\ref{eq:phi_trajectory}),
the intercepts and slopes from this determination are in good 
agreement with each other.
Whereas the intercepts agree very well with DL, the slopes of the
present measurement are clearly  smaller than that of the DL trajectory,
which was determined from $pp$ elastic scattering data. It would therefore
seem that this simple Pomeron trajectory is not universal.

\section{Summary}
Elastic and proton-dissociative photoproduction of \rhoz, \phiz\ and \jpsi\
mesons have been investigated at an average photon-proton
center-of-mass energy of 94 GeV and for values of $-t$ up to 3~GeV$^2$.
The proton-dissociative event sample was limited to values for the
mass of the dissociated proton system below 7 GeV.
The differential cross section \dsdt, for each of the vector
mesons in each of these two processes, has been measured.
The following features are common to  these reactions:
\begin{itemize}
\item 
the ratio of the differential cross sections for the elastic and
proton-dissociative reactions drops rapidly
from a value of about 1 at $-t$ = 0.4 GeV$^2$ to a value $\le$~0.1 for
$-t>$ 1 GeV$^2$;
\item
parameterization of the differential cross section as a single
exponential, $\dsdt \propto \exp{(bt)}$, for values of $-t$ exceeding 
0.5~GeV$^2$, yields a difference in the exponential slopes, $b$, for the
elastic and proton-dissociative reactions of about 3.5~GeV$^{-2}$ 
for each of the vector mesons;
\item
the measured ratio of the elastic and proton-dissociative
differential cross sections is  similar to that measured in
$pp$ elastic and single-dissociative scattering and
is consistent with the hypothesis of
Regge factorization.  
\end{itemize}

The analysis of the decay-angle distributions for pion-pair
photoproduction
in the $\rho^0$ mass region indicates a small deviation from SCHC
giving non-zero single- and double-flip amplitudes.

A comparison of the measured differential cross sections, $d\sigma/dt$, 
for the process $\gamma p \to V N$ with QCD models shows that the 
perturbative part of the calculations 
for \rhoz\ and \phiz\ production~\cite{ginzburg} 
at the $-t$ values covered in this analysis is well below the data. 
However, the perturbative QCD prediction~\cite{bartels} compares 
satisfactorily to the \jpsi\ data for values of $-t$ as low as 1~GeV$^2$.

The ratio  $\phi$ : $\rho^0$ of the elastic cross sections increases
with $-t$ and approaches 2 : 9 at $-t\approx 1$~GeV$^2$. The ratio 
$J/\psi$ : $\rho^0$ of the proton-dissociative cross sections  
increases with $-t$, but even at $-t =$ 2.2 GeV$^2$ is still
much lower than the value of 8 : 9 expected for a flavor-independent
production mechanism. 

The Pomeron trajectory was determined using elastic production of
$\phi$ and $\rho^0$ mesons
by studying the $W$ dependence of $\dsdt$ at fixed $t$
values, together with lower $W$ data from
other experiments. The resulting trajectories are:
\begin{itemize}
\item
$\gamma p \to \rho^0 p$ : \ \ \ $\alphapom(t)=(1.096\pm 0.021)+(0.125\pm
0.038)t$;
\item 
$\gamma p \to \phi p$ : \ \ \ $\alphapom(t)=(1.081\pm 0.010)+(0.158\pm
0.028)t$.
\end{itemize}
The values obtained for $\alphapom(0)$ are in good agreement with
those of DL~\cite{dl}.  However, the slopes, $\alphappom$, are
significantly lower than the value found in $p p$ elastic scattering.

In conclusion, the results presented in this analysis suggest that
even for the highest $-t$ values studied here ($-t \sim$ 2--3
GeV$^2$), the variable $t$ cannot be consistently treated as a hard
scale in perturbative QCD.

\section*{Acknowledgments}
We thank the DESY Directorate for their strong support and
encouragement. The remarkable achievements of the HERA machine group
were essential for the successful completion of this work and are
gratefully appreciated. We are grateful to
J.R. Cudell, D.Yu. Ivanov and M. W\"usthoff
for providing the theoretical calculations.
It is also a pleasure to thank L. Frankfurt
and M. Strikman for useful discussions.

%%%%%%%%%%%%%%%%%%%%%%%%%%%%%%%%%%%%%%%%%%%%%%%%%%%%%%%%%%%%%%%%%%%%%%%%%%%%%%%
\def\sp#1 {\parbox{0cm}{\rule{0cm}{#1cm}}}

\begin{table}[htb]
\begin{center}
\begin{tabular}{|c|c|} \hline
\multicolumn{2}{|c|}{\sp{0.6} ZEUS 1995 \hspace*{1cm} $\gamma p \ra \rhoz N$}  \\ \hline
\sp{0.6} $  -t $(GeV$^2$) & d$\sigma/$d$t$($\mu$b$/$GeV$^2$) \\ \hline
\sp{0.6} $  0.353 $&$ 8.8 \pm 0.7 _{-1.6}^{+1.3} \ _{-0.9}^{+1.0}$ \\ \hline  
\sp{0.6} $  0.449 $&$ 5.14 \pm 0.46 _{-0.76}^{+0.60}\ _{-0.62}^{+0.69}$ \\ \hline
\sp{0.6} $  0.548 $&$ 3.57 \pm 0.34 _{-0.30}^{+0.45}\ _{-0.55}^{+0.64}$ \\ \hline 
\sp{0.6} $  0.647 $&$ 2.68 \pm 0.28  _{-0.41}^{+0.46}\ _{-0.43}^{+0.55}$ \\ \hline
\sp{0.6} $  0.747 $&$ 2.27 \pm 0.26  _{-0.68}^{+0.47}\ _{-0.40}^{+0.52}$ \\ \hline
\sp{0.6} $  0.848 $&$ 1.62 \pm 0.20  _{-0.44}^{+0.18}\ _{-0.32}^{+0.43}$ \\ \hline
\sp{0.6} $  0.949 $&$ 1.49 \pm 0.19  _{-0.25}^{+0.20}\ _{-0.34}^{+0.49}$ \\ \hline
\sp{0.6} $  1.093 $&$ 1.07 \pm 0.12  _{-0.17}^{+0.23}\ _{-0.27}^{+0.38}$ \\ \hline
\sp{0.6} $  1.334 $&$ 0.53 \pm 0.06 _{-0.07}^{+0.10}\ _{-0.16}^{+0.26}$ \\ \hline
\sp{0.6} $  1.635 $&$ 0.26 \pm 0.06 _{-0.05}^{+0.10}\ _{-0.11}^{+0.17}$ \\ \hline
\sp{0.6} $  1.937 $&$ 0.22 \pm 0.04 _{-0.05}^{+0.04}\ _{-0.09}^{+0.18}$ \\ \hline
\sp{0.6} $  2.273 $&$ 0.064 \pm 0.018 _{-0.018}^{+0.017}\ _{-0.036}^{+0.061}$ \\ \hline
\sp{0.6} $  2.711 $&$ 0.039 \pm 0.010 _{-0.012}^{+0.012}\ _{-0.019}^{+0.049}$ \\ \hline
\end{tabular}
\end{center}
\caption{The differential cross sections, \dsdt\ , for proton-dissociative 
\rhoz \ photoproduction for $\langle W\rangle=94$~GeV
and $M_N^2<0.1W^2$. 
Statistical, systematic and uncertainties due 
to the modeling of the proton-dissociation process are given separately.
The normalization error of 15\% is not included.}
\end{table}

\begin{table}[h]
\begin{center}
\begin{tabular}{|c|c|} \hline
\multicolumn{2}{|c|}{\sp{0.6} ZEUS 1995 \hspace*{1cm} $\gamma p \ra \rhoz p$}  \\ \hline
\sp{0.6} $  -t $(GeV$^2$) & d$\sigma/$d$t$($\mu$b$/$GeV$^2$) \\ \hline
\sp{0.6} $  0.350 $&$ 5.7 \pm 0.4 _{-1.1}^{+1.3} \ _{-0.1}^{+0.1}$ \\ \hline  
\sp{0.6} $  0.448 $&$ 2.99 \pm 0.21 _{-0.58}^{+0.62}\ _{-0.06}^{+0.04}$ \\ \hline
\sp{0.6} $  0.546 $&$ 1.24 \pm 0.15 _{-0.31}^{+0.26}\ _{-0.14}^{+0.07}$ \\ \hline 
\sp{0.6} $  0.645 $&$ 0.79 \pm 0.08  _{-0.17}^{+0.16}\ _{-0.04}^{+0.03}$ \\ \hline
\sp{0.6} $  0.744 $&$ 0.50 \pm 0.06  _{-0.10}^{+0.21}\ _{-0.04}^{+0.03}$ \\ \hline
\sp{0.6} $  0.847 $&$ 0.23 \pm 0.04  _{-0.05}^{+0.13}\ _{-0.03}^{+0.03}$ \\ \hline
\sp{0.6} $  0.945 $&$ 0.183 \pm 0.033  _{-0.032}^{+0.041}\ _{-0.024}^{+0.015}$ \\ \hline
\sp{0.6} $  1.085 $&$ 0.053 \pm 0.012  _{-0.016}^{+0.011}\ _{-0.015}^{+0.013}$ \\ \hline
\sp{0.6} $  1.314 $&$ 0.019 \pm 0.006 _{-0.005}^{+0.017}\ _{-0.007}^{+0.006}$ \\ \hline
\sp{0.6} $  1.617 $&$ 0.006 \pm 0.004 _{-0.009}^{+0.002}\ _{-0.004}^{+0.004}$ \\ \hline
\end{tabular}
\end{center}
\caption{The differential cross sections, \dsdt\ , for elastic 
\rhoz \ photoproduction  for $\langle W\rangle=94$~GeV.
Statistical, systematic and uncertainties due 
to the modeling of the proton-dissociation process are given separately.
The normalization error of 15\% is not included.}
\end{table}

\begin{table}[htb]
\begin{center}
\begin{tabular}{|c|c|} \hline
\multicolumn{2}{|c|}{\sp{0.6} ZEUS 1995}  \\ \hline
\sp{0.6} $-t$(GeV$^2$) & d$\sigma/$d$t(\gamma p \ra \rhoz p) \ \ / \ \ $d$\sigma/$d$t(\gamma p \ra \rhoz
N)$\\ \hline
\sp{0.6} $  0.351 $&$ 0.66 \pm 0.07 _{-0.22}^{+0.07}\ _{-0.07}^{+0.06}$ \\ \hline  
\sp{0.6} $  0.449 $&$ 0.59 \pm 0.07 _{-0.10}^{+0.04}\ _{-0.07}^{+0.07}$ \\ \hline
\sp{0.6} $  0.547 $&$ 0.35 \pm 0.06 _{-0.07}^{+0.04}\ _{-0.07}^{+0.06}$ \\ \hline 
\sp{0.6} $  0.646 $&$ 0.30 \pm 0.05 _{-0.06}^{+0.06}\ _{-0.05}^{+0.05}$ \\ \hline
\sp{0.6} $  0.745 $&$ 0.23 \pm 0.04 _{-0.03}^{+0.03}\ _{-0.04}^{+0.04}$ \\ \hline
\sp{0.6} $  0.847 $&$ 0.14 \pm 0.04 _{-0.02}^{+0.04}\ _{-0.03}^{+0.03}$ \\ \hline
\sp{0.6} $  0.947 $&$ 0.13 \pm 0.04 _{-0.02}^{+0.07}\ _{-0.03}^{+0.02}$ \\ \hline
\sp{0.6} $  1.089 $&$ 0.05 \pm 0.02 _{-0.02}^{+0.05}\ _{-0.01}^{+0.01}$ \\ \hline
\sp{0.6} $  1.324 $&$ 0.04 \pm 0.02 _{-0.04}^{+0.07}\ _{-0.01}^{+0.01}$ \\ \hline
\sp{0.6} $  1.626 $&$ 0.03 \pm 0.02 _{-0.07}^{+0.08}\ _{-0.01}^{+0.02}$ \\ \hline
\end{tabular}
\end{center}
\caption{The ratio of the elastic to the proton dissociative cross
sections for \rhoz \ photoproduction for $\langle W\rangle=94$~GeV 
 and $M_N^2<0.1W^2$.
Statistical, systematic and uncertainties due 
to the modeling of the proton-dissociation process are given separately.
}
\end{table}

\begin{table}[htb]
\begin{center}
\begin{tabular}{|c|c|c|c|c|} \hline
\multicolumn{5}{|c|}{\sp{0.6} ZEUS 1995 \hspace*{1cm} 
$\gamma p \ra \rhoz N$}  \\ \hline
\sp{0.6} $t$ interval (GeV$^2$)& $\langle -t\rangle$(GeV$^2$) &
$\rzfzz$ & 
$\mbox{Re}[\rzfpz]$ & 
$\rzfpm$ \\ \hline
\sp{0.6} $0.30<-t<0.45$ & $0.37$ & $-0.14\pm 0.34 ^{+0.27}_{-0.20}$ & 
$-0.04\pm 0.19^{+0.10}_{-0.26}$ & $-0.05\pm 0.19^{+0.22}_{-0.11}$  \\ \hline
\sp{0.6} $0.45<-t<0.80$ & $0.60$ & $0.15\pm 0.10 ^{+0.07}_{-0.08} $ & 
$0.07\pm 0.04^{+0.06}_{-0.05}$  & $-0.04\pm 0.10^{+0.10}_{-0.13}$  \\ \hline 
\sp{0.6} $0.80<-t<1.40$ & $1.04$ & $0.01\pm 0.08 ^{+0.11}_{-0.05} $ & 
$0.05\pm 0.04^{+0.03}_{-0.03}$  & $-0.22\pm 0.06^{+0.05}_{-0.06}$  \\ \hline
\sp{0.6} $1.40<-t<3.00$ & $1.80$ & $-0.07\pm 0.10 ^{+0.11}_{-0.25}$ & 
$0.06\pm 0.04^{+0.08}_{-0.04}$  & $-0.14\pm 0.08^{+0.06}_{-0.07}$ \\ \hline  
\end{tabular}
\end{center}
\caption{The spin density matrix elements, \rzfzz, $\mbox{Re}[\rzfpz]$
and \rzfpz \
in four $t$ intervals for the proton-dissociative sample.
The data cover the kinematic range $85<W<105$~GeV and $0.55<\Mpp<1.2$~GeV.
Statistical and systematic uncertainties are given separately.}
\end{table}

\begin{table}[htb]
\begin{center}
\begin{tabular}{|c|c|c|c|c|} \hline
\multicolumn{5}{|c|}{\sp{0.6} ZEUS 1995 \hspace*{1cm} 
$\gamma p \ra \rhoz p$}  \\ \hline
\sp{0.6} $t$ interval (GeV$^2$)& $\langle -t\rangle$(GeV$^2$) &
$\rzfzz$ & 
$\mbox{Re}[\rzfpz]$ & 
$\rzfpm$ \\ \hline
\sp{0.6} $0.30<-t<0.45$ & $0.35$ & $0.03\pm 0.17 ^{+0.16}_{-0.31}$ & 
$0.14\pm 0.06^{+0.04}_{-0.09}$ & $-0.15\pm 0.10^{+0.16}_{-0.05}$  \\ \hline
\sp{0.6} $0.45<-t<0.80$ & $0.57$ & $0.09\pm 0.09 ^{+0.07}_{-0.08} $ & 
$0.08\pm 0.04^{+0.05}_{-0.03}$  & $-0.12\pm 0.07^{+0.04}_{-0.10}$  \\ \hline 
\sp{0.6} $0.80<-t<1.40$ & $0.97$ & $0.07\pm 0.15 ^{+0.12}_{-0.15} $ & 
$0.08\pm 0.06^{+0.09}_{-0.04}$  & $-0.12\pm 0.12^{+0.12}_{-0.14}$  \\ \hline
\end{tabular}
\end{center}
\caption{The spin density matrix elements, \rzfzz, $\mbox{Re}[\rzfpz]$
and \rzfpz \
in three $t$ intervals for the elastic sample.
The data cover the kinematic range $85<W<105$~GeV and $0.55<\Mpp<1.2$~GeV.
Statistical and systematic uncertainties are given separately.}
\end{table}

\begin{table}[htb]
\begin{center}
\begin{tabular}{|c|c|} \hline
\multicolumn{2}{|c|}{\sp{0.6} ZEUS 1995 \hspace*{1cm} $\gamma p \ra \phi N$}  \\ \hline
\sp{0.6} $ -t $(GeV$^2$) & d$\sigma/$d$t$($\mu$b$/$GeV$^2$) \\ \hline
\sp{0.6} $  0.470 $&$ 0.87 \pm 0.25 _{-0.21}^{+0.24}\ _{-0.15}^{+0.17}$ \\ \hline
\sp{0.6} $  0.638 $&$ 0.39 \pm 0.09 _{-0.12}^{+0.12}\ _{-0.09}^{+0.08}$ \\ \hline
\sp{0.6} $  0.901 $&$ 0.235 \pm 0.042 _{-0.041}^{+0.063}\ _{-0.051}^{+0.074}$ \\ \hline
\sp{0.6} $  1.274 $&$ 0.124 \pm 0.027 _{-0.026}^{+0.022}\ _{-0.039}^{+0.064}$ \\ \hline
\sp{0.6} $  1.969 $&$ 0.019 \pm 0.005 _{-0.003}^{+0.005}\ _{-0.009}^{+0.020}$ \\ \hline
\end{tabular}
\end{center}
\caption{The differential cross sections, \dsdt\ , for proton-dissociative 
$\phi$ photoproduction for $\langle W\rangle=94$~GeV
and $M_N^2<0.1W^2$.
Statistical, systematic and uncertainties due 
to the modeling of the proton-dissociation process are given separately.
The normalization error of 15\% is not included.}
\end{table}

\begin{table}[htb]
\begin{center}
\begin{tabular}{|c|c|} \hline
\multicolumn{2}{|c|}{\sp{0.6} ZEUS 1995 \hspace*{1cm} $\gamma p \ra \phi p$}  \\ \hline
\sp{0.6} $  -t $(GeV$^2$) & d$\sigma/$d$t$($\mu$b$/$GeV$^2$) \\ \hline
\sp{0.6} $  0.462 $&$ 0.333 \pm 0.070 _{-0.047}^{+0.094}\ _{-0.011}^{+0.015}$ \\ \hline
\sp{0.6} $  0.628 $&$ 0.168 \pm 0.027 _{-0.028}^{+0.046}\ _{-0.008}^{+0.008}$ \\ \hline
\sp{0.6} $  0.859 $&$ 0.027 \pm 0.007 _{-0.007}^{+0.011}\ _{-0.004}^{+0.003}$ \\ \hline
\sp{0.6} $  1.217 $&$ 0.004 \pm 0.003 _{-0.002}^{+0.004}\ _{-0.002}^{+0.002}$ \\ \hline
\end{tabular}
\end{center}
\caption{The differential cross sections, \dsdt\ , for elastic
$\phi$ photoproduction for $\langle W\rangle=94$~GeV.
Statistical, systematic and uncertainties due 
to the modeling of the proton-dissociation process are given separately.
The normalization error of 15\% is not included.}
\end{table}

\begin{table}[htb]
\begin{center}
\begin{tabular}{|c|c|} \hline
\multicolumn{2}{|c|}{\sp{0.6} ZEUS 1995}  \\ \hline
\sp{0.6} $-t$(GeV$^2$) & d$\sigma/$d$t(\gamma p \ra \phi p) \ \ / \ \ $d$\sigma/$d$t(\gamma p \ra
\phi N)$\\ \hline
\sp{0.6} $  0.466 $&$ 0.40 \pm 0.16 _{-0.12}^{+0.19}\ _{-0.07}^{+0.08}$ \\ \hline
\sp{0.6} $  0.633 $&$ 0.45 \pm 0.14 _{-0.15}^{+0.27}\ _{-0.07}^{+0.09}$ \\ \hline
\sp{0.6} $  0.880 $&$ 0.13 \pm 0.05 _{-0.05}^{+0.06}\ _{-0.03}^{+0.02}$ \\ \hline
\sp{0.6} $  1.245 $&$ 0.04 \pm 0.04 _{-0.02}^{+0.05}\ _{-0.01}^{+0.01}$ \\ \hline
\end{tabular}
\end{center}
\caption{The ratio of the elastic to the proton dissociative cross
sections for $\phi$ photoproduction for $\langle W\rangle=94$~GeV
and $M_N^2<0.1W^2$.
Statistical, systematic and uncertainties due 
to the modeling of the proton-dissociation process are given separately.
}
\end{table}

\begin{table}[htb]
\begin{center}
\begin{tabular}{|c|c|} \hline
\multicolumn{2}{|c|}{\sp{0.6} ZEUS 1995 \hspace*{1cm} $\gamma p \ra J/\psi N$}  \\ \hline
\sp{0.6} $  -t $(GeV$^2$) & d$\sigma/$d$t$($\mu$b$/$GeV$^2$) \\ \hline
\sp{0.6} $  0.242 $&$ 0.077 \pm 0.032 _{-0.021}^{+0.026}\ _{-0.008}^{+0.009}$ \\ \hline
\sp{0.6} $  0.625 $&$ 0.021 \pm 0.015 _{-0.006}^{+0.009}\ _{-0.004}^{+0.009}$ \\ \hline
\sp{0.6} $  1.180 $&$ 0.046 \pm 0.021 _{-0.016}^{+0.023}\ _{-0.016}^{+0.045}$ \\ \hline
\sp{0.6} $  2.169 $&$ 0.013 \pm 0.007 _{-0.005}^{+0.006}\ _{-0.007}^{+0.023}$ \\ \hline
\end{tabular}
\end{center}
\caption{The differential cross sections, \dsdt\ , for proton-dissociative
$J/\psi$ photoproduction for $\langle W\rangle=94$~GeV
and $M_N^2<0.1W^2$.
Statistical, systematic and uncertainties due 
to the modeling of the proton-dissociation process are given separately.
The normalization error of 15\% is not included.}
\end{table}

\begin{table}[htb]
\begin{center}
\begin{tabular}{|c|c|} \hline
\multicolumn{2}{|c|}{\sp{0.6} ZEUS 1995 \hspace*{1cm} $\gamma p \ra J/\psi p$}  \\ \hline
\sp{0.6} $  -t$(GeV$^2$) & d$\sigma/$d$t$($\mu$b$/$GeV$^2$) \\ \hline
\sp{0.6} $  0.219 $&$ 0.096 \pm 0.022 _{-0.020}^{+0.027}\ _{-0.006}^{+0.004}$ \\ \hline
\sp{0.6} $  0.567 $&$ 0.020 \pm 0.008 _{-0.005}^{+0.005}\ _{-0.003}^{+0.002}$ \\ \hline
\sp{0.6} $  1.084 $&$ 0.007 \pm 0.005 _{-0.003}^{+0.005}\ _{-0.003}^{+0.003}$ \\ \hline
\end{tabular}
\end{center}
\caption{The differential cross sections, \dsdt\ , for elastic
$J/\psi$ photoproduction for $\langle W\rangle=94$~GeV.
Statistical, systematic and uncertainties due 
to the modeling of the proton-dissociation process are given separately.
The normalization error of 15\% is not included.}
\end{table}

\begin{table}[htb]
\begin{center}
\begin{tabular}{|c|c|} \hline
\multicolumn{2}{|c|}{\sp{0.6} ZEUS 1995}  \\ \hline
\sp{0.6} $-t$(GeV$^2$) & d$\sigma/$d$t(\gamma p \ra J/\psi p) \ \ / \ \ $d$\sigma/$d$t(\gamma p \ra
J/\psi N)$\\ \hline
\sp{0.6} $  0.231 $&$ 1.31 \pm 0.70 _{-0.42}^{+0.50}\ _{-0.14}^{+0.18}$ \\ \hline
\sp{0.6} $  0.596 $&$ 1.08 \pm 0.97 _{-0.39}^{+0.63}\ _{-0.33}^{+0.32}$ \\ \hline
\sp{0.6} $  1.132 $&$ 0.19 \pm 0.19 _{-0.07}^{+0.20}\ _{-0.08}^{+0.02}$ \\ \hline
\end{tabular}
\end{center}
\caption{The ratio of the elastic to the proton dissociative cross
sections for $J/\psi$ photoproduction for $\langle W\rangle=94$~GeV 
 and $M_N^2<0.1W^2$.
Statistical, systematic and uncertainties due 
to the modeling of the proton-dissociation process are given separately.
}
\end{table}

\begin{table}[htb]
\begin{center}
\begin{tabular}{|c|c|} \hline
\multicolumn{2}{|c|}{\sp{0.6} ZEUS 1995 \hspace*{1cm} $\gamma p \ra \rhoz N$}  \\ \hline
\sp{0.6} $  -t $(GeV$^2$) & d$\sigma/$d$t$($\mu$b$/$GeV$^2$) \\ \hline
\sp{0.6} $  0.352 $&$ 4.69 \pm 0.36 \pm 0.90 \pm 0.16$ \\ \hline  
\sp{0.6} $  0.449 $&$ 2.61 \pm 0.23 \pm 0.42 \pm 0.15$ \\ \hline
\sp{0.6} $  0.547 $&$ 1.71 \pm 0.16 \pm 0.24 \pm 0.13$ \\ \hline 
\sp{0.6} $  0.647 $&$ 1.20 \pm 0.13 \pm 0.23 \pm 0.07$ \\ \hline
\sp{0.6} $  0.746 $&$ 0.96 \pm 0.11 \pm 0.27 \pm 0.05$ \\ \hline
\sp{0.6} $  0.846 $&$ 0.68 \pm 0.08 \pm 0.16 \pm 0.03$ \\ \hline
\sp{0.6} $  0.949 $&$ 0.60 \pm 0.08 \pm 0.11 \pm 0.03$ \\ \hline
\sp{0.6} $  1.093 $&$ 0.39 \pm 0.04 \pm 0.08 \pm 0.03$ \\ \hline
\sp{0.6} $  1.333 $&$ 0.191 \pm 0.019 \pm 0.036 \pm 0.014$ \\ \hline
\sp{0.6} $  1.638 $&$ 0.089 \pm 0.020 \pm 0.032 \pm 0.018$ \\ \hline
\sp{0.6} $  1.936 $&$ 0.071 \pm 0.012 \pm 0.019 \pm 0.010$ \\ \hline
\sp{0.6} $  2.281 $&$ 0.021 \pm 0.006 \pm 0.008 \pm 0.007$ \\ \hline
\sp{0.6} $  2.714 $&$ 0.012 \pm 0.003 \pm 0.004 \pm 0.002$ \\ \hline
\end{tabular}
\end{center}
\caption{The differential cross sections, \dsdt\ , for proton-dissociative 
\rhoz \ photoproduction for $\langle W\rangle=94$~GeV
and $M_N^2<0.05W^2$. 
Statistical, systematic and uncertainties due 
to the modeling of the proton-dissociation process are given separately.
The normalization error of 15\% is not included.}
\end{table}

\begin{table}[htb]
\begin{center}
\begin{tabular}{|c|c|} \hline
\multicolumn{2}{|c|}{\sp{0.6} ZEUS 1995 \hspace*{1cm} $\gamma p \ra \phi N$}  \\ \hline
\sp{0.6} $ -t $(GeV$^2$) & d$\sigma/$d$t$($\mu$b$/$GeV$^2$) \\ \hline
\sp{0.6} $  0.467 $&$ 0.43 \pm 0.12 \pm 0.16 \pm 0.05$ \\ \hline
\sp{0.6} $  0.639 $&$ 0.19 \pm 0.04 \pm 0.07 \pm 0.02$ \\ \hline
\sp{0.6} $  0.895 $&$ 0.096 \pm 0.017 \pm 0.026 \pm 0.004$ \\ \hline
\sp{0.6} $  1.264 $&$ 0.046 \pm 0.010 \pm 0.014 \pm 0.005$ \\ \hline
\sp{0.6} $  1.977 $&$ 0.007 \pm 0.002 \pm 0.002 \pm 0.001$ \\ \hline
\end{tabular}
\end{center}
\caption{The differential cross sections, \dsdt\ , for proton-dissociative 
$\phi$ photoproduction for $\langle W\rangle=94$~GeV
and $M_N^2<0.05W^2$.
Statistical, systematic and uncertainties due 
to the modeling of the proton-dissociation process are given separately.
The normalization error of 15\% is not included.}
\end{table}

\begin{table}[htb]
\begin{center}
\begin{tabular}{|c|c|} \hline
\multicolumn{2}{|c|}{\sp{0.6} ZEUS 1995 \hspace*{1cm} $\gamma p \ra J/\psi N$}  \\ \hline
\sp{0.6} $  -t $(GeV$^2$) & d$\sigma/$d$t$($\mu$b$/$GeV$^2$) \\ \hline
\sp{0.6} $  0.236 $&$ 0.066 \pm 0.027 \pm 0.033 \pm 0.004$ \\ \hline
\sp{0.6} $  0.616 $&$ 0.013 \pm 0.009 \pm 0.010 \pm 0.001$ \\ \hline
\sp{0.6} $  1.170 $&$ 0.023 \pm 0.010 \pm 0.012 \pm 0.001$ \\ \hline
\sp{0.6} $  2.152 $&$ 0.006 \pm 0.003 \pm 0.003 \pm 0.001$ \\ \hline
\end{tabular}
\end{center}
\caption{The differential cross sections, \dsdt\ , for proton-dissociative
$J/\psi$ photoproduction for $\langle W\rangle=94$~GeV
and $M_N^2<0.05W^2$.
Statistical, systematic and uncertainties due 
to the modeling of the proton-dissociation process are given separately.
The normalization error of 15\% is not included.}
\end{table}
\begin{table}[htb]
\begin{center}
\begin{tabular}{|c|c|} \hline
\multicolumn{2}{|c|}{\sp{0.6} ZEUS 1995 }  \\ \hline
\sp{0.6} $-t$(GeV$^2$) & d$\sigma/$d$t(\gamma p \ra \phi N) \ \ / \ \ $d$\sigma/$d$t(\gamma p \ra \rhoz N)$ \\ \hline
\sp{0.6} $  0.492 $&$ 0.145 \pm 0.028 _{-0.052}^{+0.037}$ \\ \hline
\sp{0.6} $  1.007 $&$ 0.149 \pm 0.021 _{-0.027}^{+0.018}$ \\ \hline
\sp{0.6} $  2.047 $&$ 0.172 \pm 0.056 _{-0.048}^{+0.090}$ \\ \hline
\end{tabular}
\end{center}
\caption{The ratio of the cross sections \dsdt\ for \phiz \ to \rhoz\ for
proton-dissociative photoproduction for $\langle W\rangle=94$~GeV and 
$M_N^2<0.1W^2$.
Statistical and systematic  uncertainties are given
separately.}
\end{table}

\begin{table}[htb]
\begin{center}
\begin{tabular}{|c|c|} \hline
\multicolumn{2}{|c|}{\sp{0.6} ZEUS 1995 }  \\ \hline
\sp{0.6} $-t$(GeV$^2$) & d$\sigma/$d$t(\gamma p \ra \phi p) \ \ / \ \ $d$\sigma/$d$t(\gamma p \ra \rhoz
p)$ \\ \hline
\sp{0.6} $  0.464 $&$ 0.155 \pm 0.023 _{-0.025}^{+0.046}$ \\ \hline
\sp{0.6} $  0.865 $&$ 0.225 \pm 0.047 _{-0.048}^{+0.132}$ \\ \hline
\end{tabular}
\end{center}
\caption{The ratio of the cross sections \dsdt\ for \phiz \ to \rhoz\ for
elastic photoproduction for $\langle W\rangle=94$~GeV.
Statistical and systematic  uncertainties are given
separately.}
\end{table}

\begin{table}[htb]
\begin{center}
\begin{tabular}{|c|c|} \hline
\multicolumn{2}{|c|}{\sp{0.6} ZEUS 1995 }  \\ \hline
\sp{0.6} $-t$(GeV$^2$) & d$\sigma/$d$t(\gamma p \ra J/\psi N) \ \ / \ \ $d$\sigma/$d$t(\gamma p \ra \rhoz N)$ \\ \hline
\sp{0.6} $  0.499 $&$ 0.006 \pm 0.004 _{-0.001}^{+0.002}$ \\ \hline
\sp{0.6} $  1.044 $&$ 0.032 \pm 0.014 _{-0.012}^{+0.005}$ \\ \hline
\sp{0.6} $  2.134 $&$ 0.097 \pm 0.058 _{-0.029}^{+0.020}$ \\ \hline
\end{tabular}
\end{center}
\caption{The ratio of the cross sections \dsdt\ for \jpsi \ to \rhoz\ for
proton-dissociative photoproduction for $\langle W\rangle=94$~GeV
and $M_N^2<0.1W^2$.
Statistical and systematic  uncertainties are given
separately.}
\end{table}

\begin{table}[htb]
\begin{center}
\begin{tabular}{|c|c|} \hline
\multicolumn{2}{|c|}{\sp{0.6} ZEUS 1995 }  \\ \hline
\sp{0.6} $-t$(GeV$^2$) & d$\sigma/$d$t(\gamma p \ra J/\psi p) \ \ / \ \ $d$\sigma/$d$t(\gamma p \ra \rhoz
p)$ \\ \hline
\sp{0.6} $  0.475 $&$ 0.015 \pm 0.005 _{-0.003}^{+0.003}$ \\ \hline
\sp{0.6} $  0.901 $&$ 0.045 \pm 0.030 _{-0.017}^{+0.030}$ \\ \hline
\end{tabular}
\end{center}
\caption{The ratio of the cross sections \dsdt\ for \jpsi \ to \rhoz\ for
elastic photoproduction for $\langle W\rangle=94$~GeV. 
Statistical and systematic uncertainties are given
separately.}
\end{table}

%%%%%%%%%%%%%%%%%%%%%%%%%%%%%%%%%%%%%%%%%%%%%%%%%%%%%%%%%%%%%%%%%%%%%%%%%%%%%%


\begin{thebibliography}{99}
\addcontentsline{toc}{chapter}{Bibliografy}
\bibitem{jim} For a recent review see e.g.
J.A.~Crittenden, {\it Exclusive Production of Neutral Vector Mesons at
the Electron-Proton Collider HERA,} {\rm Springer Tracts in Modern
Physics}, Volume ~140 (Springer, Berlin Heidelberg, 1997).
\bibitem{abramowicz-caldwell} See H. Abramowicz and A. Caldwell, 
DESY 98--192 (1998), accepted by Rev. Mod. Phys, and references therein.
\bibitem{large-t} L. Frankfurt and M. Strikman, \prl{63}{1989}{1914}; 
\newline A.H. Mueller and  W-K. Tang, \pl{B284}{1992}{123}.
\bibitem{halfms} H. Abramowicz, L. Frankfurt, M. Strikman, \shep{11}{1997}{51}.
\bibitem{vdm} J.J.Sakurai, \ap{11}{1960}{1}; \\
J.J Sakurai, \prl{22}{1969}{981}.
\bibitem{bauer} See e.g. T.H. Bauer et al., \rmp{50}{1978}{261};
\ Erratum \rmp{51}{1979}{407}.
\bibitem{chapin} T.J. Chapin et al., \prev{D31}{1985}{17}. 
\bibitem{zrho-94} ZEUS Collab., J. Breitweg et al., \ejp{C2}{1998}{247}.
\bibitem{bfgms} 
S.J.~Brodsky et al., \prev{D50}{1994}{3134}.
\bibitem{ryskin} 
M.G.~Ryskin, \zp{C57}{1993}{89};\\
M.G.~Ryskin, R.G.~Roberts, A.D.~Martin and E.M.~Levin, \zp{C76}{1997}{231}.
\bibitem{bartels}J. Bartels, J.R. Forshaw, H. Lotter and M. W\"usthoff,
\pl{B375}{1996}{301}. 
\bibitem{ivanov} D.Yu. Ivanov, \prev{D53}{1996}{3564}.
\bibitem{czerniak} V.L. Chernyak and A.R. Zhitnitsky, \prep{112}{1984}{173}.
\bibitem{ginzburg} I.F. Ginzburg and D.Yu Ivanov, \prev{D54}{1996}{5523}.
\bibitem{collins} P.D.B. Collins, {\it An Introduction to Regge Theory
and High-Energy Physics}, Cambridge University Press, Cambridge,
England, 1977.
\bibitem{jl-slope} G.A. Jaroszkiewicz and P.V. Landshoff, \prev{D10}{1974}{170}.
\bibitem{collins-slope} P.D.B. Collins, F.D. Gault and A. Martin, 
\np{B80}{1974}{135}.
\bibitem{giacomelli} G. Giacomelli, \prep{23C}{1976}{123}.
\bibitem{burq-slope} J.P. Burq et al., \pl{B109}{1982}{124};
\ \ \np{B217}{1983}{285}.
\bibitem{dl-slope} A. Donnachie and P.V. Landshoff, \np{B231}{1984}{189}.
\bibitem{noshrink} A. Levy, \pl{B424}{1998}{191}.
\bibitem{gl} E. Gotsman and A. Levy, \prev{D13}{1976}{3036}. 
\bibitem{barber-83} D.P. Barber et al., \zp{C12}{1982}{1}. 
\bibitem{detector} ZEUS Collab., M. Derrick et al., 
\pl{B297}{1992}{404};\ \ \pl{B303}{1993}{183};\ \ The\  ZEUS\  Detector,
\ Status\ Report\ 1993, \ ed. \ U. \ Holm.
\bibitem{ctd} N. Harnew et al., \nim{A279}{1989}{290};\\
B. Foster et al., {\it Nucl. Phys. {\bf B}, Proc-Suppl.} 
{\bf B32} (1993) 181; \\
B. Foster et al., \nim{A338}{1994}{254}.
\bibitem{CAL} A. Andresen et al., \nim{A309}{1991}{101};\\
A. Caldwell et al., \nim{A321}{1992}{356};\\
A. Bernstein et al., \nim{A336}{1993}{23}. 
\bibitem{lumi} D. Kisielewska et al.,
DESY-HERA 85-25 (1985);\\
J.Andruszk\'ow et al., DESY 92-066 (1992).
\bibitem{dipsi} M. Arneodo, L. Lamberti, R. Ryskin, \cpc{100}{1997}{195}.
\bibitem{epsoft} M. Kasprzak, PhD thesis, Warsaw University, 
DESY F35D-96-16(1996).
\bibitem{chlm} CHLM Collab., M. Albrow et al., \np{B108}{1976}{1}.
\bibitem{LA_thesis} L. Adamczyk, PhD thesis, Academy of Mining and
  Metallurgy, Cracow (1999).
\bibitem{jpsi_intercept} ZEUS Collab., {\it Study of vector meson 
production at large $-t$ at HERA and determination of the Pomeron
trajectory}, paper 788 submitted to ICHEP98, Vancouver, 23-29 July, 1998.
\bibitem{cool} R.L. Cool et al., \prl{48}{1982}{1451}.
\bibitem{kurek} K. Kurek, DESY 96-209 (1996).
\bibitem{soding} P. S\"oding, \pl{19}{1966}{702}.
\bibitem{rs} M. Ross and L. Stodolsky, \prev{149}{1966}{1172}.
\bibitem{ref:pdg} Particle Data Group: Review of Particle Properties, 
\prev{D50}{1994}{1}.
\bibitem{zvm95} ZEUS Collab., J. Breitweg et al., \ejp{C6}{1999}{603}. 
\bibitem{zrho-lps} ZEUS Collab., M. Derrick et al., \zp{C73}{1997}{253}.
\bibitem{h1-rho-deltab} H1 Collab., C. Adloff et al., \zp{C75}{1997}{607}.
\bibitem{wolf} K. Schilling, P. Seyboth and G. Wolf, \np{B15}{1970}{397};\\
K. Schilling and G. Wolf, \np{B61}{1973}{381}.
\bibitem{schc} F.J. Gilman et al., \pl{31B}{1970}{387}.
\bibitem{ballam-73} J. Ballam et al., \prev{D7}{1973}{3150}.
\bibitem{zbpc} ZEUS Collab., J. Breitweg et al., DESY 99-102 (1999).
\bibitem{h1schc} H1 Collab., C. Adloff et al., DESY 99-010 (1999).
\bibitem{zphi-94} ZEUS Collab., M. Derrick et al., \pl{B377}{1996}{259}.
\bibitem{zphi/rho} ZEUS Collab., M. Derrick et al,
  \pl{B380}{1996}{220}.
\bibitem{zjpsi-94} ZEUS Collab., M. Derrick et al., \zp{C75}{1997}{215}.
\bibitem{h1phi/rho} H1 Collab., C. Adloff et al., \zp{C75}{1997}{607}.
\bibitem{h1jpsi/rho} H1 Collab., S. Aid et al., \np{B468}{1996}{3}.
\bibitem{ref:isabelle}I. Royen and J.-R. Cudell, \np{B545}{1999}{505}. 
\bibitem{gribov-fact} V.N. Gribov and L.Ya Pomeranchuk, \prl{8}{1962}{343}.
\bibitem{kokkedee} J.J.J. Kokkedee, {\it The quark model},
W.A. Benjamin, Inc., New York, USA, 1969.
\bibitem{omega} D. Aston et al., \np{B209}{1982}{56}.
\bibitem{h1rho-94} H1 Collab., S. Aid et al., \np{B463}{1996}{3}.
\bibitem{dl} A. Donnachie and P.V. Landshoff, \pl{B296}{1992}{227}.
\bibitem{fruend} P.G.O. Freund, \nc{48}{1967}{541}.
\bibitem{behrend-78} H.J. Behrend et al., \np{B144}{1978}{22}.
\bibitem{mcclellan-71} G. McClellan et al., \prl{26}{1971}{1593}.
\bibitem{anderson-70} R. Anderson et al., \prev{D1}{1970}{27}.
\bibitem{busenitz-89} J. Busenitz et al., \prev{D40}{1989}{1}.
\end{thebibliography}
\end{document}